\documentclass[a4paper]{spie}  
\usepackage[pdftex]{graphicx} 
\usepackage{comment}

\usepackage{url} 
\usepackage{multirow} 
\usepackage{dirtytalk} 
\usepackage{amssymb} 

\title{Concept Design of Low Frequency Telescope for CMB B-mode Polarization satellite LiteBIRD} 

\author[1,2,3]{Y.~Sekimoto}
\author[4]{P.A.R.~Ade}
\author[5]{A.~Adler}
\author[6]{E.~Allys}
\author[7]{K.~Arnold}
\author[8]{D.~Auguste}
\author[9]{J.~Aumont}
\author[10]{R.~Aurlien}
\author[11]{J.~Austermann}
\author[12]{C.~Baccigalupi}
% ...
\affil[1]{Japan Aerospace Exploration Agency (JAXA), Institute of Space and Astronautical Science (ISAS), Sagamihara, Kanagawa 252-5210, Japan}
\affil[2]{The University of Tokyo, Department of Astronomy, Tokyo 113-0033, Japan}
\affil[3]{High Energy Accelerator Research Organization (KEK), Tsukuba, Ibaraki 305-0801, Japan}
\affil[4]{Cardiff University, School of Physics and Astronomy, Cardiff CF10 3XQ, UK}
\affil[5]{Stockholm University}
\affil[6]{Laboratoire de Physique de l’$\acute{\rm E}$cole Normale Sup$\acute{\rm e}$rieure, ENS, Universit$\acute{\rm e}$ PSL, CNRS, Sorbonne Universit$\acute{\rm e}$, Universit$\acute{\rm e}$ de Paris, 75005 Paris, France}
\affil[7]{University of California, San Diego, Department of Physics, San Diego, CA 92093-0424, USA}
\affil[8]{Universit\'e Paris-Saclay, CNRS/IN2P3, IJCLab, 91405 Orsay, France}
\affil[9]{IRAP, Universit$\acute{\rm e}$ de Toulouse, CNRS, CNES, UPS, (Toulouse), France}
\affil[10]{University of Oslo, Institute of Theoretical Astrophysics, NO-0315 Oslo, Norway}
\affil[11]{National Institute of Standards and Technology (NIST), Boulder, Colorado 80305, USA}
\affil[12]{International School for Advanced Studies (SISSA), Via Bonomea 265, 34136, Trieste, Italy}
% ...

\authorinfo{Send correspondence to Y. Sekimoto, E-mail: sekimoto.yutaro@jaxa.jp}

\begin{document} 
  \maketitle 


\begin{abstract}
 LiteBIRD has been selected as JAXA's strategic large mission in the 2020s,
 to observe the cosmic microwave background (CMB) $B$-mode polarization over the full sky at large angular scales. 
The challenges of LiteBIRD are the wide field-of-view (FoV) and broadband capabilities of millimeter-wave polarization measurements, which are derived from the system requirements. 
The possible paths of stray light increase with a wider FoV and the far sidelobe knowledge of $-56$ dB is a challenging optical requirement.
A crossed-Dragone configuration was chosen for the low frequency telescope (LFT : 34--161 GHz), one of LiteBIRD's onboard telescopes.
It has a wide field-of-view ($18^\circ \times 9^\circ$) with an aperture of 400 mm in diameter, corresponding to an angular resolution of about 30 arcminutes around 100 GHz.
The focal ratio f/3.0 and the crossing angle of the optical axes of 90$^\circ$ are chosen after an extensive study of the stray light. The primary and secondary reflectors have rectangular shapes with serrations to reduce the diffraction pattern from the edges of the mirrors. 
The reflectors and structure are made of aluminum to proportionally contract from warm down to the operating temperature at $5\,$K. 
A 1/4 scaled model of the LFT has been developed to validate the wide field-of-view design and to demonstrate the reduced far sidelobes.
A polarization modulation unit (PMU), realized with a half-wave plate (HWP) is placed in front of the aperture stop, the entrance pupil of this system.
A large focal plane with approximately 1000 AlMn TES detectors and frequency multiplexing SQUID amplifiers is cooled to 100 mK. The lens and sinuous antennas have broadband capability. 
Performance specifications of the LFT and an outline of the proposed verification plan are presented.
\end{abstract}



\section{INTRODUCTION}
\label{sec:intro}  
LiteBIRD, the Lite (Light) satellite for the study of $B$-mode polarization and Inflation from cosmic background
Radiation Detection,  observes the cosmic microwave background (CMB) polarization over the full sky at large angular scales \cite{Hazumi2012,Sekimoto2018,Hazumi2020}. 
    Cosmological inflation predicts primordial gravitational waves, which imprinted large-scale curl ($B$-mode) patterns on the CMB polarization map \cite{PhysRevLett.78.2054,PhysRevLett.78.2058,PhysRevD.55.1830,PhysRevD.55.7368}.
Measurements of the CMB $B$-mode signals are known as the best probe to detect the primordial gravitational waves and to measure the inflation energy. 
The scientific objective of LiteBIRD is to test major inflationary models \cite{Kamionkowski2016}. 
The power of the $B$-modes is proportional to the tensor-to-scalar ratio, $r$. 
The current upper limit on $r$ is  $r < 0.044$\cite{Tristram2020}. 
The mission goal of LiteBIRD is to measure $r$ with a precision of $\delta r < 0.001$, which provides a crucial test of cosmic inflation. 
The required angular coverage is $2 < \ell < 200$, where $\ell$ is the multipole moment.

 LiteBIRD has been selected as JAXA's strategic large mission in the late 2020s.
It will be launched with an H3 vehicle for three years of observations at the Lagrangian point (L2) of the Earth-Sun system. 
  It is a spinning satellite with a precession angle ($\alpha$) of 45$^{\circ}$ and spin angle ($\beta$) of 50$^{\circ}$ with spin rate of 0.05 rpm and precession period of 180 minutes, which are optimized from crossing angles and revisits of previously scanned regions.
The concept design has been studied by researchers from Japan, U.S., Canada, and Europe since September 2016.  

LiteBIRD observes millimeter waves from 34$\,$GHz to 448 GHz with two instruments, LFT and MHFT\cite{Ludo2020,Lamagna2020}.
Both instruments have the same relative bandwidth of min: max frequencies = 1:5. 
LFT will explore synchrotron and CMB emission, while MHFT covers CMB emission and will also extend to higher frequencies to explore the dust contribution.
The bands in common between the two telescopes, i.e. 89--161$\,$GHz, allow reduction of systematics associated with the telescopes, and add redundancy.
A transmissive half-wave plate (HWP) for polarization modulation has a limited bandwidth, and so LiteBIRD has two instruments to cover the frequency bands.
Both instruments are operated at cryogenic temperature of $5\,$K to reduce the photon noise.
The focal plane design is based on multi-chroic TES detectors at 100$\,$mK operation \cite{Aritoki2018,Westbrook2020}.
Cryogenic chain of LiteBIRD is described by Hasebe et al. \cite{Hasebe2019} and Duval et al.\cite{Duval2020}

Challenges for LiteBIRD are wide field-of-view (FoV) and broadband capabilities of millimeter-wave polarization measurements, 
which are derived from the sensitivity specifications.
The wide FoV corresponds to a large focal plane area; a detector pixel has different spill-over or edge-taper depending on the pixel position on the focal plane.
The possible paths of stray light increase with a wider FoV.
A stable system is also required to perform the all sky survey.

LiteBIRD is currently under the conceptual study phase. It is important to define preliminary design specifications in order to make progress on the system design. The derivation of the detailed requirements and the detailed design study are moving in parallel, and affect each other iteratively. 
In this paper we introduce a list of design specifications in this phase. Based on further simulation-based studies of the error budget allocation over the entire system, the numbers we list for the design specifications may change.


\section{Overview of LFT}

\begin{figure}[hbt]
\begin{center}
   \includegraphics[width =120mm]{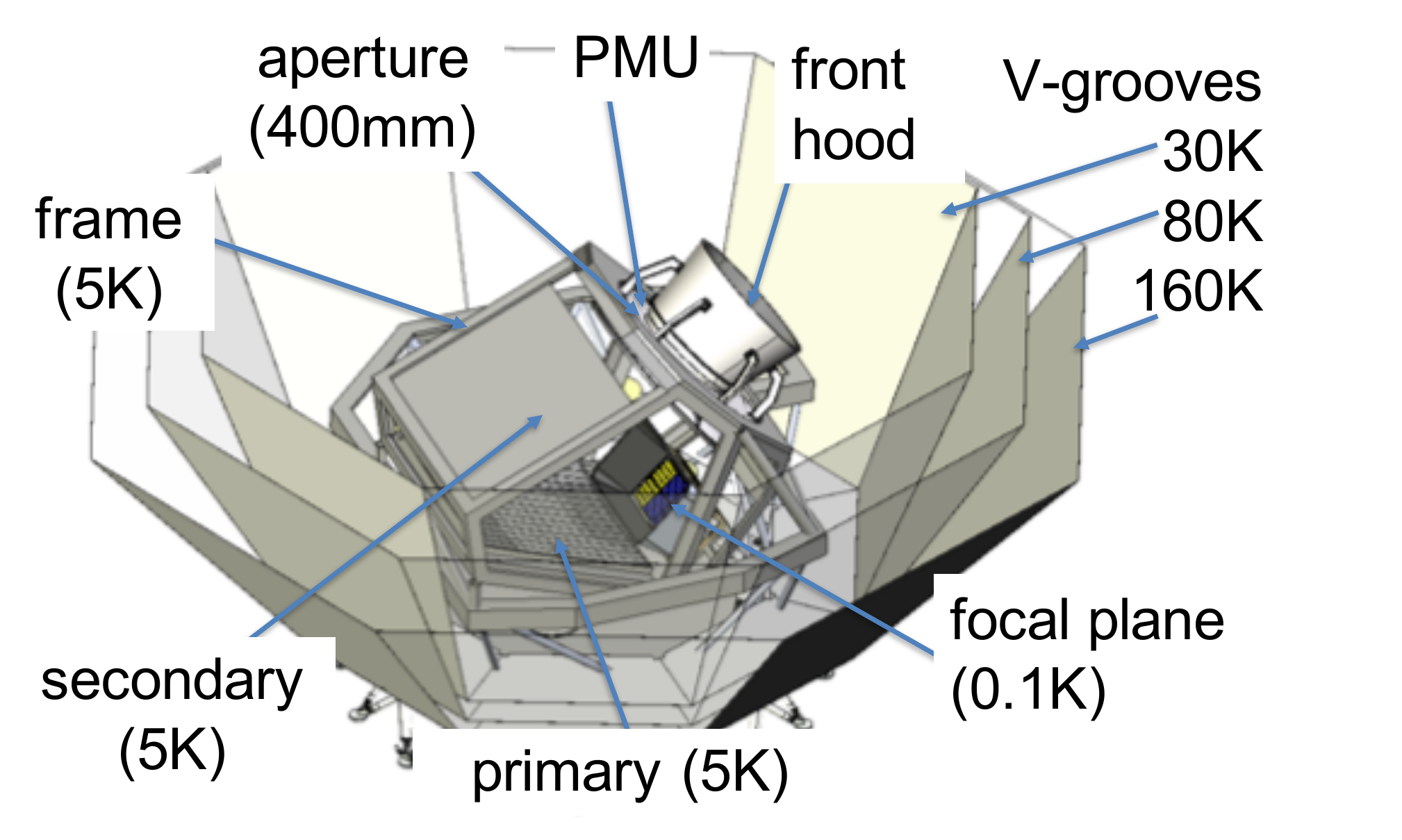}
 \end{center}
 \caption{Overview of low frequency telescope (LFT). MHFT and side panels are not shown for clarity. }
 \label{fig:LFT-overview}
\end{figure}

LFT has been designed to meet specifications described in the next section.
This section describes a brief overview of LFT before describing design details.
LFT is a wide field-of-view telescope designed to observe the CMB and synchrotron radiation in the frequencies of 34--161 GHz, as shown in Figure \ref{fig:LFT-overview}. 
The aperture diameter is 400 mm. The angular resolution is 24--71 arcminutes. 
LFT is operated at cryogenic temperature of 5$\,$K to reduce the optical loading and is surrounded by radiators called V-grooves.
The thermal design of LiteBIRD is described in Hasebe et al \cite{Hasebe2019}.
LFT has a crossed Dragone antenna made of aluminum.
A frame structure at $5\,$K supports all components: the PMU (polarization modulation unit); focal plane; primary and secondary reflectors; and absorbers.
An earlier design\cite{Sekimoto2018} has been updated.

A PMU with a transmissive HWP (half-wave plate)\cite{Sakurai2020} is mounted in front of the aperture stop. 
LFT focal plane is based on multi-chroic TES detectors at 100 mK operation \cite{Aritoki2018,Westbrook2020}.
There are interfaces with the LFT PMU and the LFT focal plane.

\section{LFT design specifications}
\label{sct:LFT-requirements}

The performance specifications for LFT are as follows.

\subsection{Frequency bands and noise}
\begin{description}
\item [Frequency coverage ] 34--161 GHz
\item [Band sensitivities]
LFT shall have the array sensitivities as tabulated in Table \ref{tbl:frequency-bands}, which shall satisfy the map-level sensitivity specifications.
The sensitivity is limited by the number of pixels, which is closely related with the field of view of the telescope.
The noise of the detector in a pixel is limited by the optical loading.

\begin{table}[htb]
    \centering
    \caption{Performance specifications of LFT. The bandwidth (BW) is (High $-$ Low)/Center frequency. }
    \begin{tabular}{|r|c|c|c|c|c|c|}
    \hline
        Center  freq. & BW & Beam fwhm & pixel dia.   & No. det & NET$_{\rm array}$  & Pol. sensitivity\\
        GHz &  & [arcmin] &  [mm] &  & [$\mu$Krts] & [$\mu$K arcmin] \\ \hline
        40& 0.30 & 70.5 & 32 & 48 & 18.5 & 37.4 \\ \hline
        50& 0.30 & 58.5 & 32 & 24 & 16.5 & 33.5 \\ \hline
        60& 0.23 & 51.1 & 32 & 48 & 10.5 & 21.3 \\ \hline
        68& 0.23 & 41.6 & 16 & 144 & 9.8 & \multirow{2}{*}{16.9} \\ 
        68& 0.23 & 47.1 & 32 & 24 & 15.7   & \\ \hline
        78& 0.23 & 36.9 & 16 & 144 & 7.7 & \multirow{2}{*}{12.1} \\ 
        78 & 0.23 & 43.8 & 32 & 48 & 9.5 &  \\ \hline
        89& 0.23 & 33.0 & 16 & 144 & 6.1 & \multirow{2}*{11.3} \\
        89 & 0.23 & 41.5 & 32 & 24 & 14.2 &  \\ \hline
        100& 0.23 & 30.2 & 16 & 144 & 5.1 & 6.6 \\ \hline
        119& 0.30 & 26.3 & 16 & 144 & 3.8 & 4.6 \\ \hline
        140& 0.30 & 23.7 & 16 & 144 & 3.6 & 4.8 \\ \hline
    \end{tabular}
    \label{tbl:frequency-bands}
\end{table}

\item [Band shape]
The frequency bandpasses are defined by a combination of superconducting band-pass filters on the wafer \cite{Aritoki2018}, and the use of quasi-optical metal-mesh filters \cite{Ade2006} in front of the focal plane to reject higher frequencies.
Lower frequencies than the defined band (red-leak) might contribute to sidelobes due to the distorted beam pattern. The red-leak is rejected only by a superconducting band-pass filter on the wafer\cite{Aritoki2018}.
Higher frequencies than the defined band (blue-leak) might contribute to noise due to far-infrared radiation.
The blue leak is rejected by both the on-chip filter and the quasi-optical metal mesh filter in front of the focal plane.

\item [1/f noise]
The knee frequency of the post-demodulation $1/f$ noise should be below $0.1\,$mHz (assuming a $0.05\,$rpm spin rate, precession angle $\alpha=45^\circ$ and spin angle $\beta =50^\circ$). 
The knee frequency of the raw $1/f$ noise should be well below 3.1 Hz ($46\,$rpm$\times 4$).
The 46 rpm value is the LFT HWP rotation rate.
In the HWP failure mode, the pair-differenced $f_{\rm knee}$ is $20\,$mHz for individual detectors and $100,$mHz for the common mode.

\item [Data loss and operational duty cycle] 
The operating life of the instruments should be long enough to perform observations for 3 years.
The system shall have an operational duty cycle of 85$\,$\% for science observations, including all downtime for cryogenic cycling, detector operation preparation, and data transfer. 
Data loss due to cosmic ray glitches should be less than 5$\,$\%.

\end{description}

\subsection{Beam}
\begin{description}
\item [Angular resolution] 
The angular resolution of each detector response should be sufficient to cover the required 
angular scales of $2< \ell < 200$, where $\ell$ is the spherical harmonic index. 
It shall have a FWHM of $80'$ or better. 
Angular resolution should be better than $30'$ at $100\,$GHz, for measuring the recombination bump, which is the prominent structure at degree scales in the $B$-mode power spectrum coming from the primordial gravitational waves. 
It shall also be better than $80'$ at $40\,$GHz, for dealing with point sources.
\item [Pointing offset knowledge]
The pointing offset knowledge should be less than $2.1'$\cite{Ishino2016,Ishino2016b}.

\item [Far sidelobe knowledge] The extended component of the far sidelobe should be known at a precision level of $-56$ dB \cite{Nagata2019,Lee2019}.
Radiation from the Galactic plane through the far sidelobes contaminates the signal and therefore the inferred power
spectrum.
The far sidelobe is currently defined as the domain located above 0.2$\,$rad.
\item [Small scale feature of sidelobe] The small-scale features of the far sidelobes should be known at a precision level of $-33\,$dB, more specifically defined by the following equation: (intensity/$0.05\,$\%)$ \times$ (diameter/$30'$)$^2$, where the diameter is the FWHM of the small-scale features due to possible optical ghosts or optical multiple reflections.
\item [Near sidelobe knowledge] The beam pattern of near sidelobes (out to 10$^\circ$ from the co-polar beam peak) should be known at a precision level of $-30\,$dB. 
Also, it should be confirmed to be consistent with its designed pattern at a precision level of 10$\,$\% or better. 

\item [Beam stability knowledge] 
The beam-shape stability over time, should be better than 0.46$\,$\% (synchronous) / $2\,$\%( random) for beam width, and better than 1.7$''$ (synchronous) / 16$''$ (random) for pointing, better than $0.086\,$\% (synchronous) / $2.7\,$\% (random) at the third flattening (often called ellipticity), 
better than $-46$ dB at sidelobes around several to $30\,$degrees \cite{Ishino2016b}. 
The time scale of the synchronous beam fluctuation is $163\,$msec for LFT in which HWP rotates by $45^{\circ}$, while "random" is a component that fluctuates randomly over time.
They correspond to \say{differential beam shape} and 
are also related to optical qualities of the instrument in the broad sense. 
Note that in the case of a perfect polarization modulator, differential beam effects are negligibly small. 
Therefore beam stability specifications are tied to imperfections of the polarization modulation system.
\end{description}

\subsection{Polarization}

\begin{description}
\item [Knowledge of polarization efficiency] 
The polarization efficiency knowledge should be better than $0.2\,$\%. 
\item [Absolute polarization angle knowledge (monopole)]	
The absolute polarization angle knowledge on the stable monopole component should be better than $2.7'$\cite{Ishino2016,Ishino2016b}. 
\item [Polarization modulation]
The modulation frequency should be $> 4 \times 0.76\,$Hz,
which assures 4 modulation (at least) during beam-size excursions of 30$'$.
The modulation frequency should be $< 4\times 4.5\,$Hz, given by an argument about the bolometer time constant.
\item [Modulation synchronous instrumental polarization knowledge]	The $4f$ synchronous instrumental polarization knowledge should be better than $0.0063\,$\%. 
\end{description}

\subsection{Gain}
\begin{description}
\item [Gain variation in time]
The gain variation in time for a single detector should be better than $10\,$\% assuming that the gain parameter is updated every 1200 sec (which corresponds to a 0.05 rpm rotation period).
The effective differential gain should be smaller than $0.0069\,$\% 
(synchronous, i.e., $163\,$msec for LFT, in which the HWP rotates by $45^\circ$) / $0.3\,$\% (random). 
\end{description}

\subsection{Other specifications}

There are other specifications.
According to the system design \cite{Sekimoto2018}, heat dissipation of LFT is limited to 4 mW, which includes the PMU and temperature control of the LFT optical components.
The minumum eigen-frequency for LFT is assumed to be 100 Hz and 50 Hz for axial and lateral axes, respectively; however, this might be optimized by a combined design with the cryo-structure of the payload module (PLM).
LFT is designed to withstand quasi-static loads of 20 g for the axial and lateral axes.
EMC/EMI specifications have been studied with simulations\cite{Tsuji2020}.

\section{Optical design}

\subsection{Antenna design}
After trade-off studies of various optical configurations among crossed-Dragone,  offset-Gregorian \cite{Tran2008}, and open-Dragone\cite{Bernacki2012,Young2018}, we concluded that the crossed-Dragone antenna is the best option for LFT because of the wide-field of view and the low cross polarization.
Multiple reflections of crossed-Dragone antennas have been described earlier\cite{Tran2010}.

\begin{table}
\centering
\caption{Optical specifications of LFT antenna}
\begin{tabular}{lc}
\hline
Aperture diameter & 400 mm \\
Field of view & $18^{\circ} \times 9^\circ$ \\
Strehl ratio &  $> 0.95$  at $161\,$GHz\\
Focal plane telecentricity &  $< 1.0^\circ$ \\
Focal ratio  & $2.9 < $ F/\# $< 3.1$ \\
PSF flattening & $< 5\,$\% \\
Cross polarization  & $< -30$ dB \\
Rotation of polarization angle across FoV & $< \pm 1.5^\circ $ \\
\hline
\end{tabular}
\label{tbl:lft-specification}
\end{table}

A crossed-Dragone antenna of LFT has been designed with anamorphic aspherical surfaces \cite{Kashima2018} to achieve the specifications listed in Table \ref{tbl:lft-specification}.
The anamorphic aspherical surface is described with the following equation 
for both the primary mirror (PM) and the secondary mirror (SM) \cite{Kashima2018}:
\begin{equation}
z_m = \frac{C_{m,x} x_m^2 + C_{m,y} y_m^2}{1 + \sqrt{1 - (1 +k_{m,x})C^2_{m,x} x_m^2 - (1 + k_{m,y}) C^2_{m,y} y_m^2}} 
+ \sum_{i=2}^5 A_{m,i} \left[ \left(1 - B_{m,i} \right) x_m^2 + \left(1 + B_{m,i} \right) y_m^2 \right]^i,
\end{equation}
where $m = $ PM, SM, $C_{m,x}$ and $C_{m,y}$ are curvatures for the $x$
and $y$ directions, $k_{m,x}$ and $k_{m,y}$ are conic constants in the x and y directions, 
and $A_{m,i}$ and $B_{m,i}$ are aspherical coefficients. 

\begin{table} [htb]
    \centering
    \caption{Optical parameters of anamorphic aspherical surfaces\cite{Kashima2018}.}
    \label{tbl:aas}
    \begin{footnotesize}
    \begin{tabular}{l|l|l|l|l|l|l|l|l}
    \hline \hline
         & $C_{m,x}$/mm$^{-1}$ & $C_{m,y}$/mm$^{-1}$ & $k_{m,x}$ & $k_{m,y}$ & $y_{m,0}$/mm & $z_{m,0}$/mm & $\theta_m$/deg. &  \\ \hline
        PM & $-1.60053\times 10^{-4}$ & $-4.71355\times 10^{-4}$ & 15.857906 &$ -5.174224$ & 0 & 696.344 & 0 &  \\ \hline
        SM & 4.05234$\times 10^{-4}$ & 5.04062$\times 10^{-4}$ & $-4.162644$ & $-1.282787$ & $-163.771$ & 346.223 & 42.45664 &  \\ \hline
        FP &  &  &  &  & 550.924 & 343.223 & 90 &  \\ \hline \hline
         & $A_{m,2}$ & $B_{m,2}$ & $A_{m,3}$ & $B_{m,3}$ & $A_{m,4}$ & $B_{m,4}$ & $A_{m,5}$ & $B_{m,5}$ \\ \hline
        PS & $-5.28\times 10^{-12}$ & $-3.31\times 10^{-1}$ & 1.63$\times 10^{-18}$ & $-0.716$ & $-2.50\times 10^{-24}$ & $-0.973$ & $-2.17\times 10^{-34}$ & 0.0929 \\ \hline
        SM & $-3.10\times 10^{-16}$ & 59.834 & 7.42$\times 10^{-18}$ & $-0.375$ & $-3.45\times 10^{-23}$ & -1.157 & $-3.89\times 10^{-31}$ & $-0.349$ \\ \hline
    \end{tabular}
        \end{footnotesize}
\end{table}

\begin{figure}[htb]
\begin{minipage}{0.5\hsize}
  \begin{center}
\includegraphics[width = 0.8\textwidth]{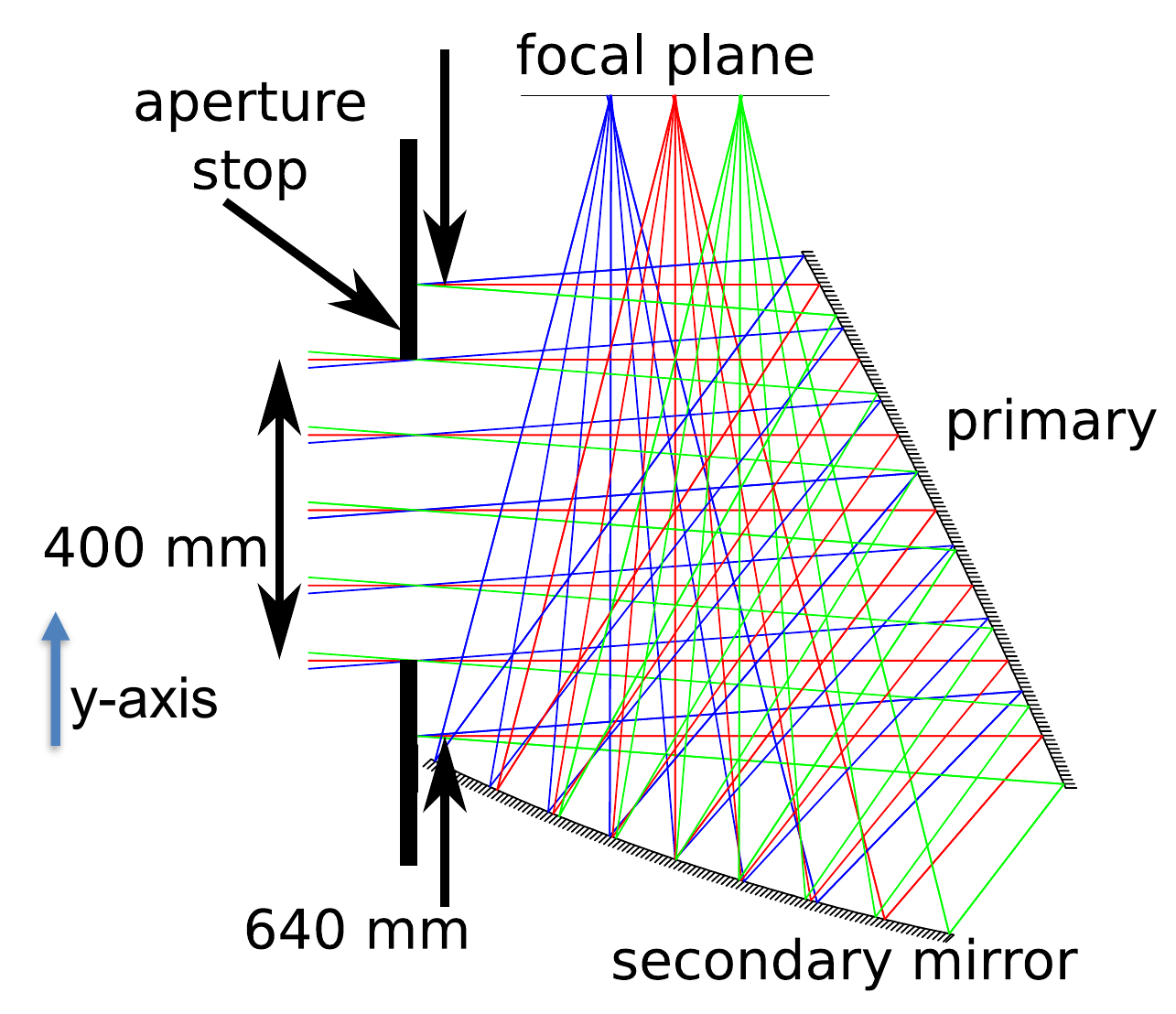}
\end{center}
 \end{minipage}
 \begin{minipage}{0.5\hsize}
  \begin{center}
  \includegraphics[width = 0.8\textwidth]{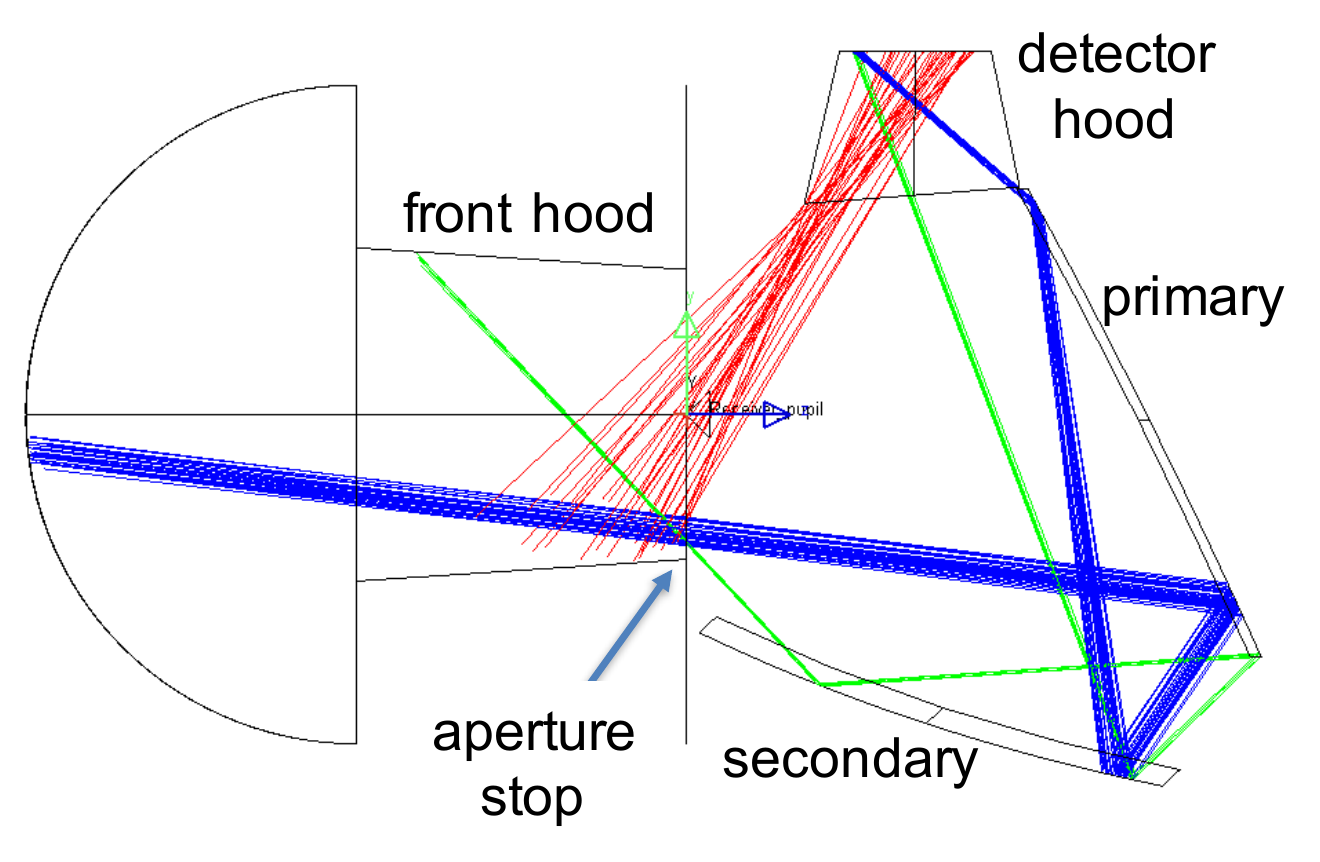}
  \end{center}
 \end{minipage}
 \caption{(Left) Ray tracing diagram of Low Frequency Telescope (LFT). Blue, Red, and Green lines show $\theta_y=+4.5^\circ$, $\theta_y=0^\circ$, $\theta_y=-4.5^\circ$, respectively. (Right) Possible stray light paths of LFT. Red lines show direct paths. Blue and green lines show triple reflections.}
\label{fig:lft-optical}
\end{figure}

A ray diagram of LFT is shown in Figure \ref{fig:lft-optical}, which has an aperture diameter of 400 mm and an FoV of 18$^\circ$$\times$ 9$^\circ$.
The aperture diameter is derived from the requirement of the angular resolution of 80$'$ at 40$\,$GHz.
The FoV corresponds to the focal plane area of $420\,$mm$\,\times 210\,$mm, which is roughly proportional to the sensitivity.
This meets the sensitivity requirement in Table \ref{tbl:frequency-bands}.

\begin{figure}[htb]
\begin{minipage}{0.5\hsize}
  \begin{center}
\includegraphics[width = 0.99\textwidth]{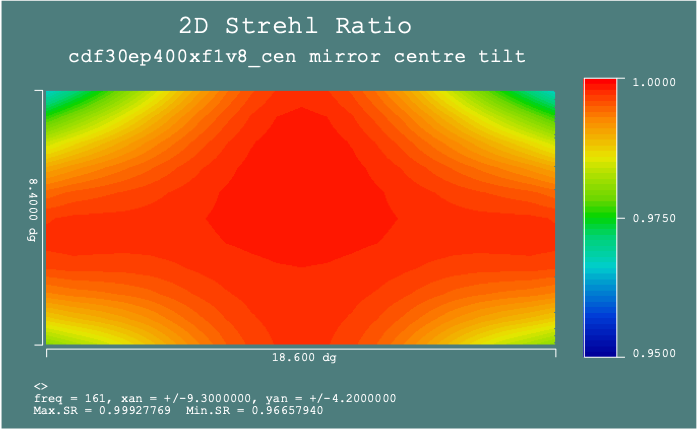}
\end{center}
 \end{minipage}
 \begin{minipage}{0.5\hsize}
  \begin{center}
  \includegraphics[width = 0.99\textwidth]{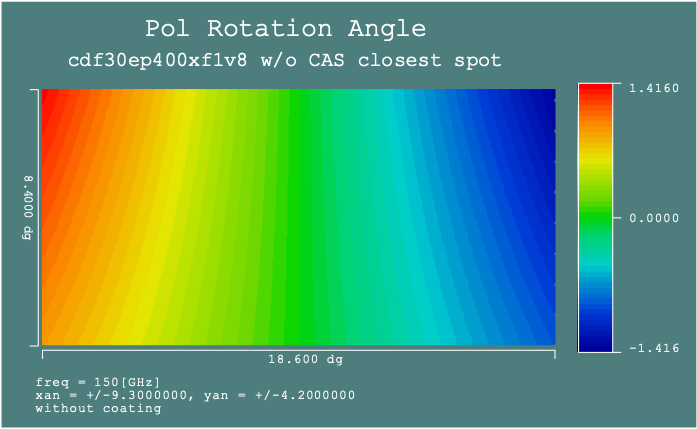}
  \end{center}
 \end{minipage}
 \vspace{1mm}
 \caption{(Left) Map of Strehl ratio of LFT antenna at $161\,$GHz. (Right) Rotation of polarization angle of $y$-axis polarization across the field of view in units of degrees. }
\label{fig:lft-strehl}
\end{figure}

Optical rays are designed to have $640\,$mm diameter at the aperture from the focal plane to keep enough edge tapers at both primary and secondary reflectors.
The Strehl ratio at $161\,$GHz is larger than 0.95, as shown in Figure~\ref{fig:lft-strehl}.
Rotation of the polarization angle for the $y$-axis polarization across the field of view is shown in Figure~\ref{fig:lft-strehl}.
The rotation is estimated to $< \pm 1.5^\circ$ according to the ray tracing simulation with a finite resistivity.
The derived optical parameters are tabulated in Table \ref{tbl:aas}.

\begin{figure}[htb]
\begin{minipage}{0.6\hsize}
  \begin{center}
\includegraphics[width = 0.8\textwidth]{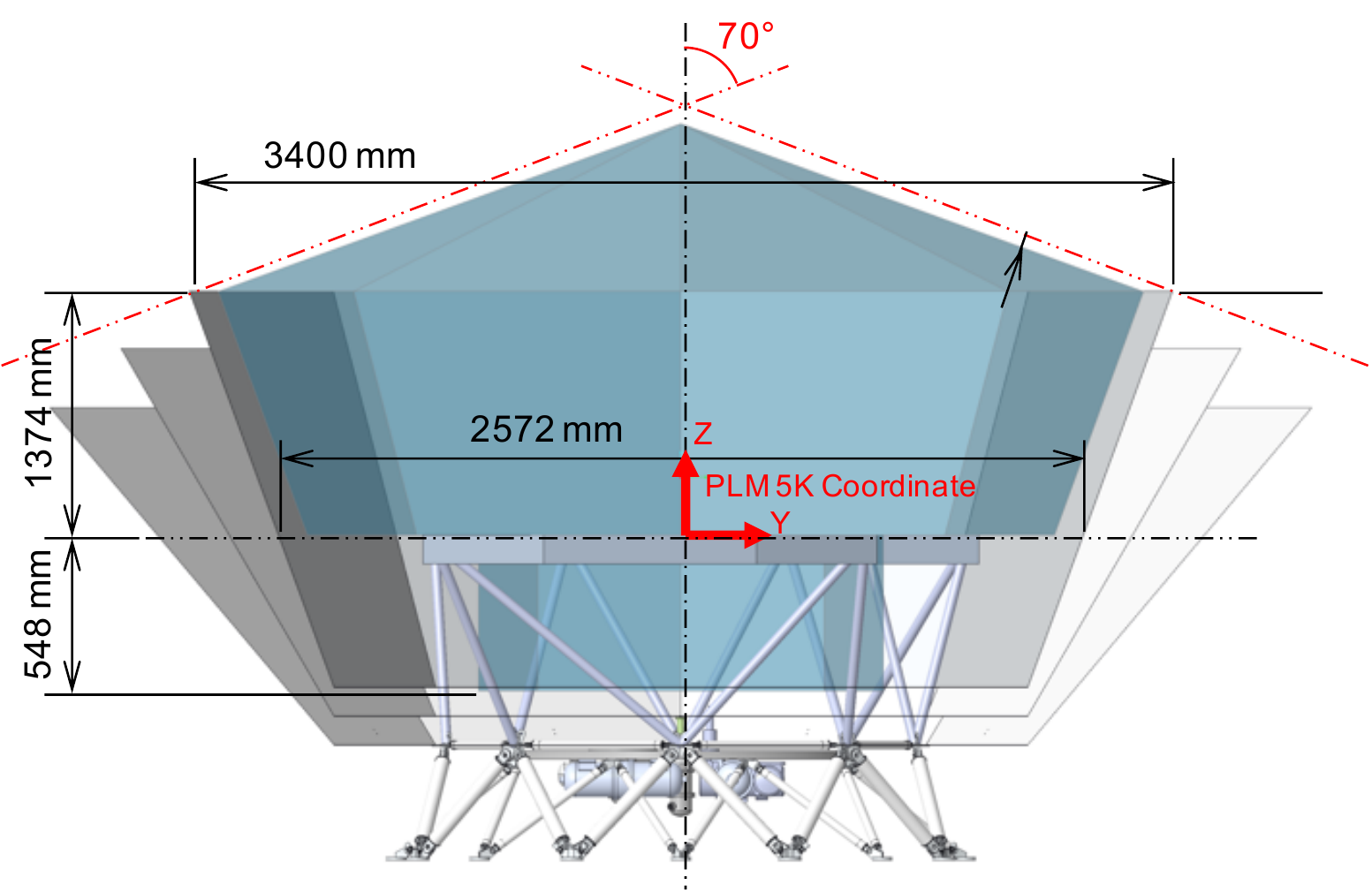}
\end{center}
 \end{minipage}
 \begin{minipage}{0.4\hsize}
  \begin{center}
  \includegraphics[width = 0.8\textwidth]{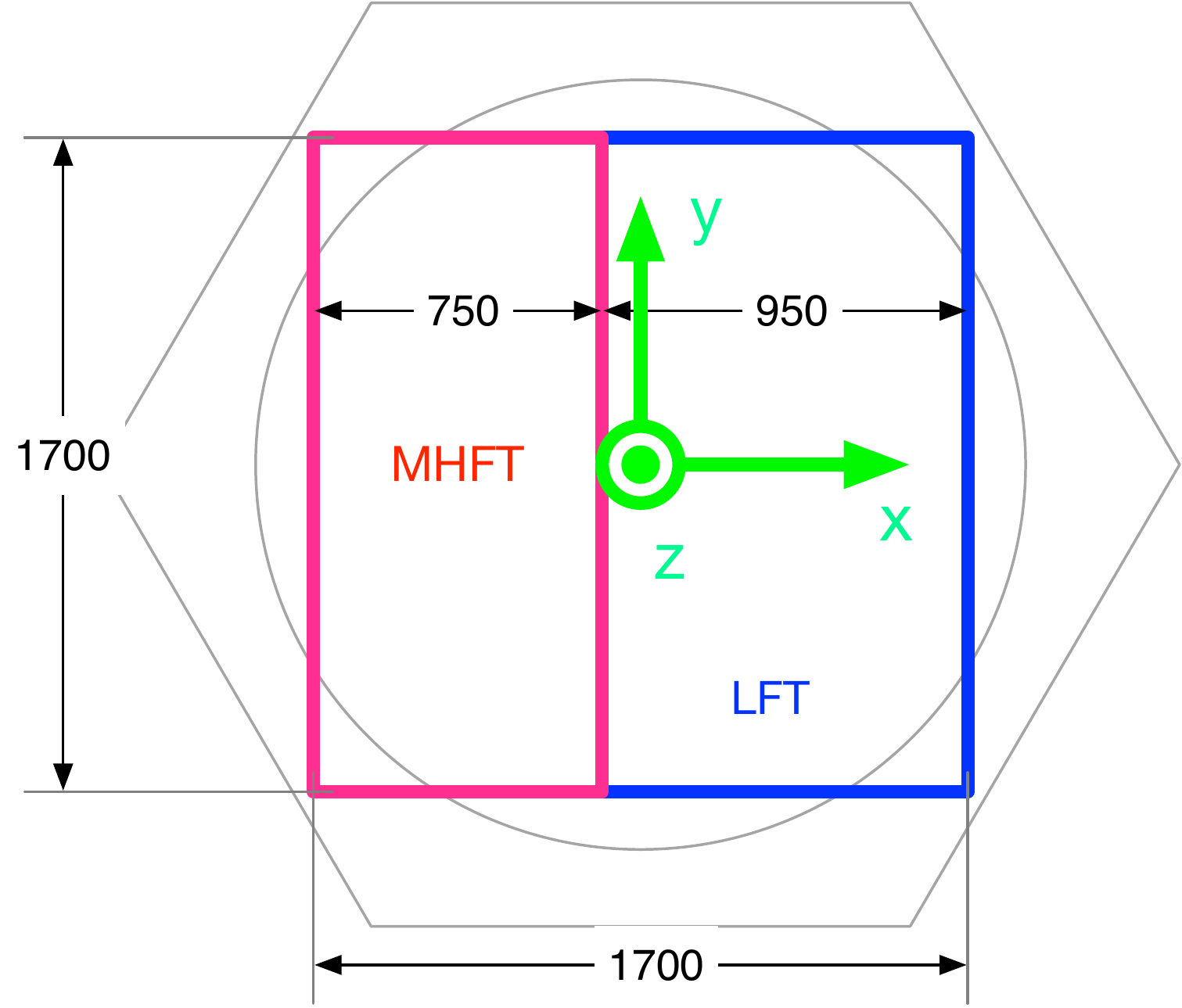}
  \end{center}
 \end{minipage}
 \caption{(Left) Usable volume of LFT and MHFT and the PLM coordinate. 
 V-grooves are also shown. The most inner V-groove is at $30\,$K. The top of the truss is the $5\,$K structural interface for LFT and MHFT.
 (Right) Allocated area of LFT and MHFT and the PLM coordinate.}
\label{fig:allocated-volume}
\end{figure}

The allocated volumes of LFT and MHFT are shown in Figure \ref{fig:allocated-volume}.
The field of view of LFT is maximized under the volume constraint.
Crossed-Dragone antennae with f/2.5, 3.0, and 3.5 are compared. The volume is roughly proportional to the f-number. Under the volume constraint, the smaller values are preferable, but, the stray light is larger. We chose f/3.0 for LFT, considering focal-plane dimensions and feed parameters.

We updated the design of a crossed-Dragone antenna reported by \cite{Kashima2018}.
The f/3.0 and the crossing angle of the optical axes of 90$^\circ$ have been chosen after an extensive study of stray light (on the right of Figure \ref{fig:lft-optical}).

\begin{figure}[htb]
\centering
  \includegraphics[width = 0.9\textwidth]{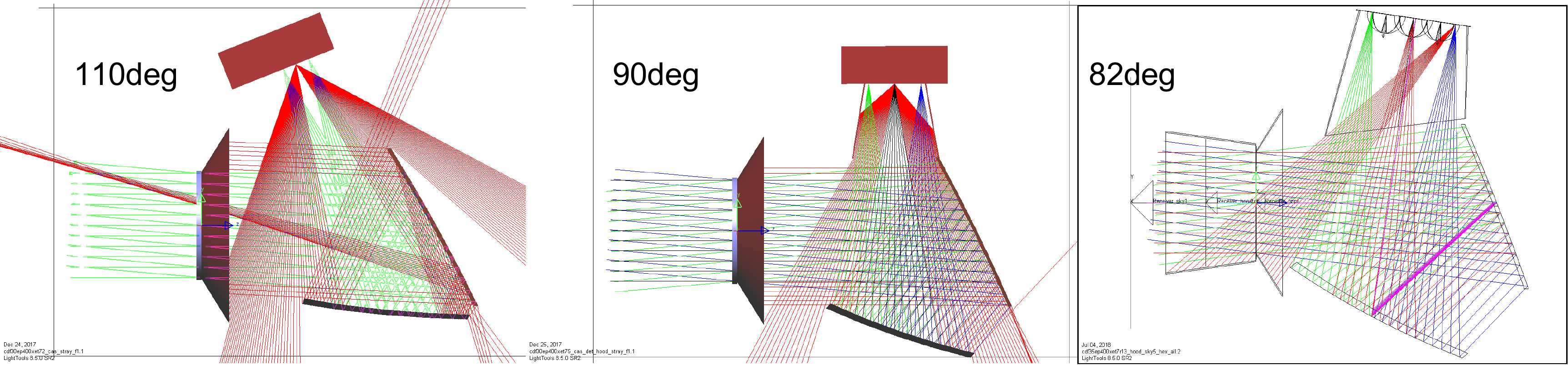}
 \caption{Stray light with the crossing angle of the optical axes of the crossed-Dragone configuration.}
\label{fig:crossing-angles}
\end{figure}

Figure \ref{fig:crossing-angles} shows the stray light with the crossing angles; 
At the crossing angle of 110$^\circ$, the direct path from the feed to the sky is small, but there are many triple reflection paths. 
At 82$^{\circ}$, there are large direct paths.
Then the 90$^\circ$ angle moderates for both the triple reflections and direct paths.

\begin{figure}[hbt]
\begin{center}
   \includegraphics[width =80mm]{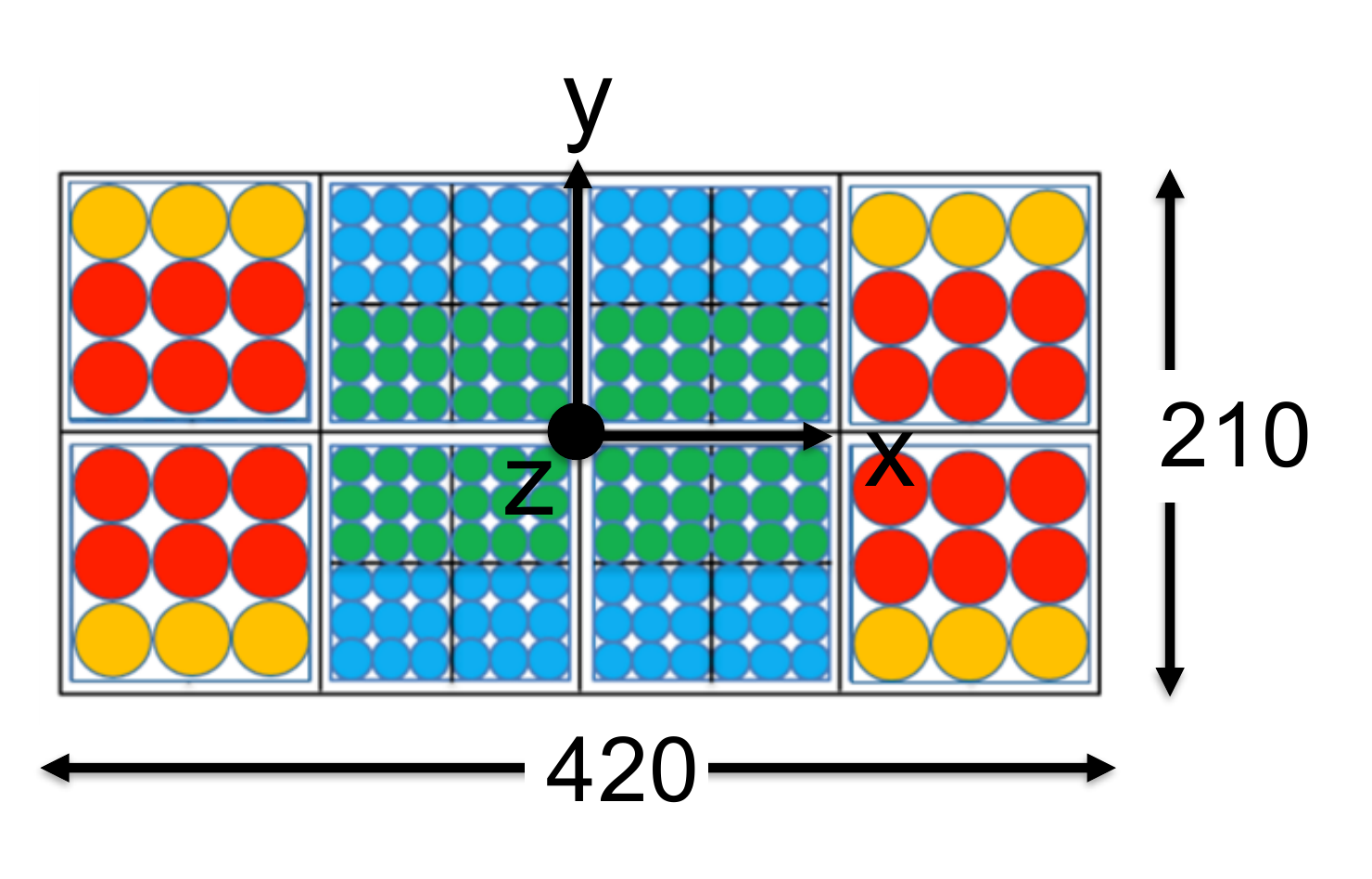}
 \end{center}
 \caption{LFT focal plane pixel arrangement. There are eight square (10 cm $\times$ 10 cm) tiles. Red,  yellow, and green, blue pixels correspond Type 1, 2, 3, and 4 of Table \ref{tbl:feed-parameters}, respectively. The LFT focal plane coordinate is shown in black arrows. The scales are shown in units of millimeters. }
 \label{fig:LFT-focal-plane-pixel}
\end{figure}

The detector hood and front hood whose height of 500 mm reduces stray light to far sidelobes as shown in Figure~\ref{fig:lft-optical}. 
The $y$-direction of the focal plane in the focal plane coordinate (Figure \ref{fig:LFT-focal-plane-pixel}) is limited by multiple reflections or stray light.
The $x$-direction is limited by the $5\,$K allocated area of LiteBIRD, as shown in Figure \ref{fig:allocated-volume}.

Primary and secondary mirrors have rectangular shapes of 835 $\times$ 795 mm and 872 $\times$ 739 mm, respectively, 
with serrations to reduce diffraction patterns from the edges of mirrors.
The mirror sizes were reduced from the previous design \cite{Sekimoto2018} because the $2\,$K cold aperture stop was removed due to limitations of the cooling capacity and then the length between the aperture and the main reflector was reduced.
The optical design is based on feed parameters as tabulated in Table \ref{tbl:feed-parameters}.


\begin{table} [htb]
    \centering
    \caption{Frequency bands and feed parameters. The bandwidth (BW) is (High $-$ Low)/Center frequency. The number (No.) of detectors is two times the number of pixels because of two orthogonal polarization detections.}
    \label{tbl:feed-parameters}
    \begin{tabular}{|c|r|c|r|r|c|c|c|c|}
    \hline
        Type & Center freq.  & BW & Low & High & Pixel dia.  & Beam waist & No. pix & No. det. \\ 
         & [GHz] &  & [GHz] & [GHz] & [mm] & radius [mm] &  &  \\ \hline
        1 & 40 & 0.30 & 34 & 46 & 32 & 11.64 & 24 & 48 \\ \hline
         & 60 & 0.23 & 53 & 67 & 32 & 11.64 & 24 & 48 \\ \hline
         & 78 & 0.23 & 69 & 87 & 32 & 11.64 & 24 & 48 \\ \hline
        2 & 50 & 0.30 & 43 & 58 & 32 & 11.64 & 12 & 24 \\ \hline
         & 68 & 0.23 & 60 & 76 & 32 & 11.64 & 12 & 24 \\ \hline
         & 89 & 0.23 & 79 & 99 & 32 & 11.64 & 12 & 24 \\ \hline
        3 & 68 & 0.23 & 60 & 76 & 16 & 5.82 & 72 & 144 \\ \hline
         & 89 & 0.23 & 79 & 99 & 16 & 5.82 & 72 & 144 \\ \hline
         & 119 & 0.30 & 101 & 137 & 16 & 5.82 & 72 & 144 \\ \hline
        4 & 78 & 0.23 & 69 & 87 & 16 & 5.82 & 72 & 144 \\ \hline
         & 100 & 0.23 & 89 & 112 & 16 & 5.82 & 72 & 144 \\ \hline
         & 140 & 0.30 & 119 & 161 & 16 & 5.82 & 72 & 144 \\ \hline
    \end{tabular}
\end{table}

\subsection{Optical simulation}
\label{sect:optical-simulation}

\begin{figure} [htb]
\begin{center}
   \includegraphics[width =160mm]{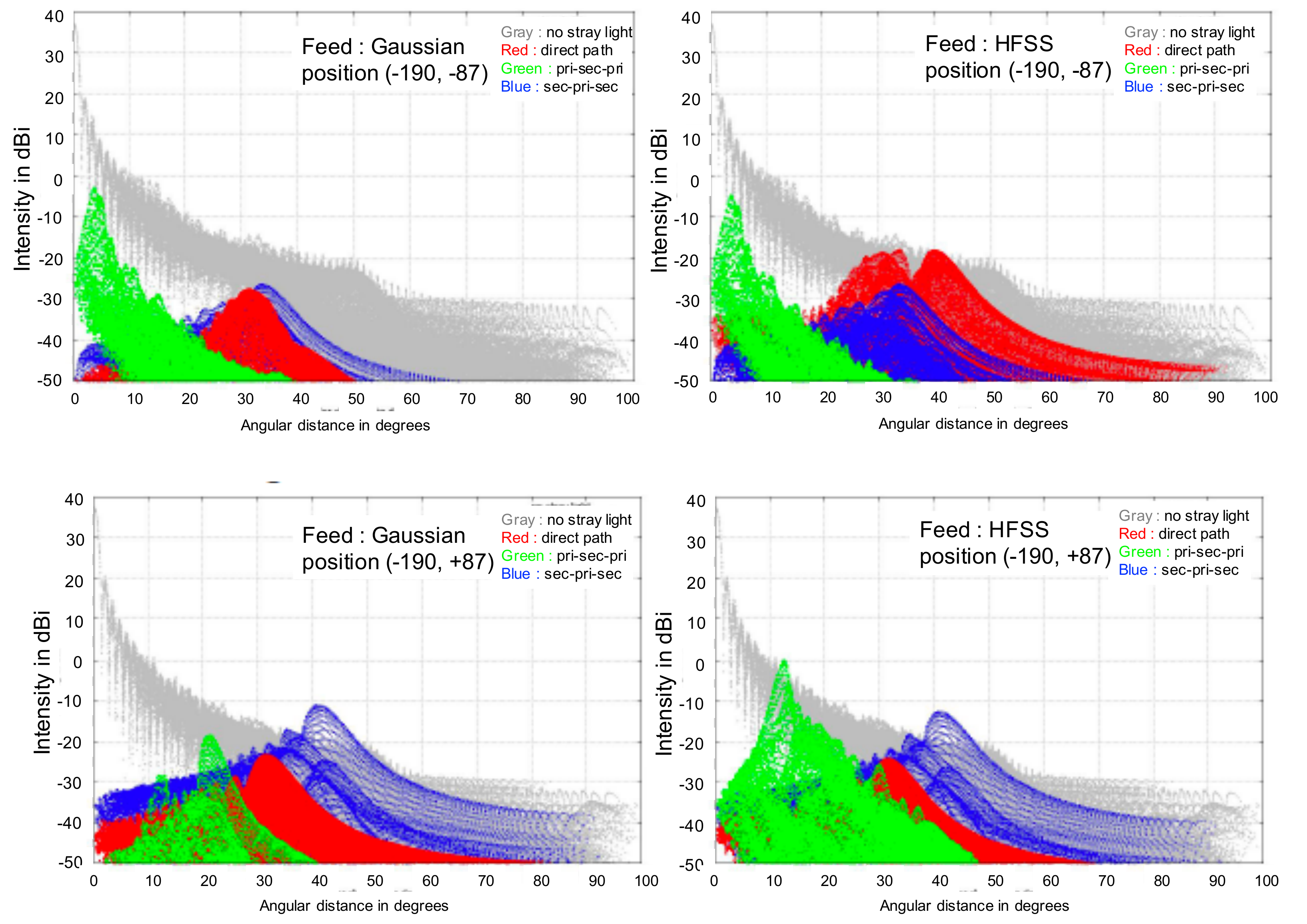}
 \end{center}
 \caption{Optical simulation of far-field beam pattern of LFT at 34$\,$GHz. Gray shows the nominal beam pattern without stray light, 
 red shows the direct path from the focal plane to sky, green shows triple reflections (feed $-$ primary $-$ secondary $-$ primary $-$ sky), and blue shows triple reflections (feed $-$ secondary $-$ primary $-$ secondary $-$ sky). (Top, Left) A pixel near the primary reflector around ($x, y$) = ($-190\,$mm, $-87\,$mm) with a Gaussian feed. (Top, Right) A pixel near the primary reflector with HFSS simulation of sinuous antenna. The feed sidelobe contributes the far sidelobe of LFT due to direct path (Red). 
 (Bottom, Left) A pixel near the aperture stop around ($x, y$) = ($-190\,$mm, $+87\,$mm) with a Gaussian feed. 
 (Top, Right) A pixel near the aperture stop with HFSS simulation of sinuous antenna. 
 The feed sidelobe contributes the far sidelobe of LFT due to triple reflections (feed $-$ primary $-$ secondary $-$ primary $-$ sky: Green). }
 \label{fig:GRASP34-imada}
\end{figure}

Physical optics simulations of LFT with GRASP10\cite{GRASP} have been studied in the same way by Imada et al.\cite{Imada2018}.
Lower frequencies make it relatively difficult to meet the far sidelobe requirement due to diffraction effects.
Figure \ref{fig:GRASP34-imada} shows the impact of the feed sidelobes.

The LFT antenna pattern assuming a Gaussian feed is shown in the left panels of Figure \ref{fig:GRASP34-imada}, while the feed simulated with HFSS\cite{HFSS} is shown in the right ones.
Upper panels show the antenna pattern of a pixel near the primary reflector, while lower ones show that of near the aperture.
It is clear that the direct path from the feed sidelobe contributes the far sidelobe of LFT at a level of $-60\,$dB.
The feed sidelobe of a pixel near the aperture contributes the point-like sidelobe due to triple reflections 
(feed $\rightarrow$ primary $\rightarrow$ secondary $\rightarrow$ primary $\rightarrow$ sky: shown in green).
Note that there are discrepancies of the feed sidelobes at a level around $-20\,$dB between the HFSS simulation and the room-temperature measurement of the sinuous/lens feed \cite{Edwards2012}.

\begin{figure}[htb]
\begin{minipage}{0.5\hsize}
\centering
\includegraphics[height = 7cm]{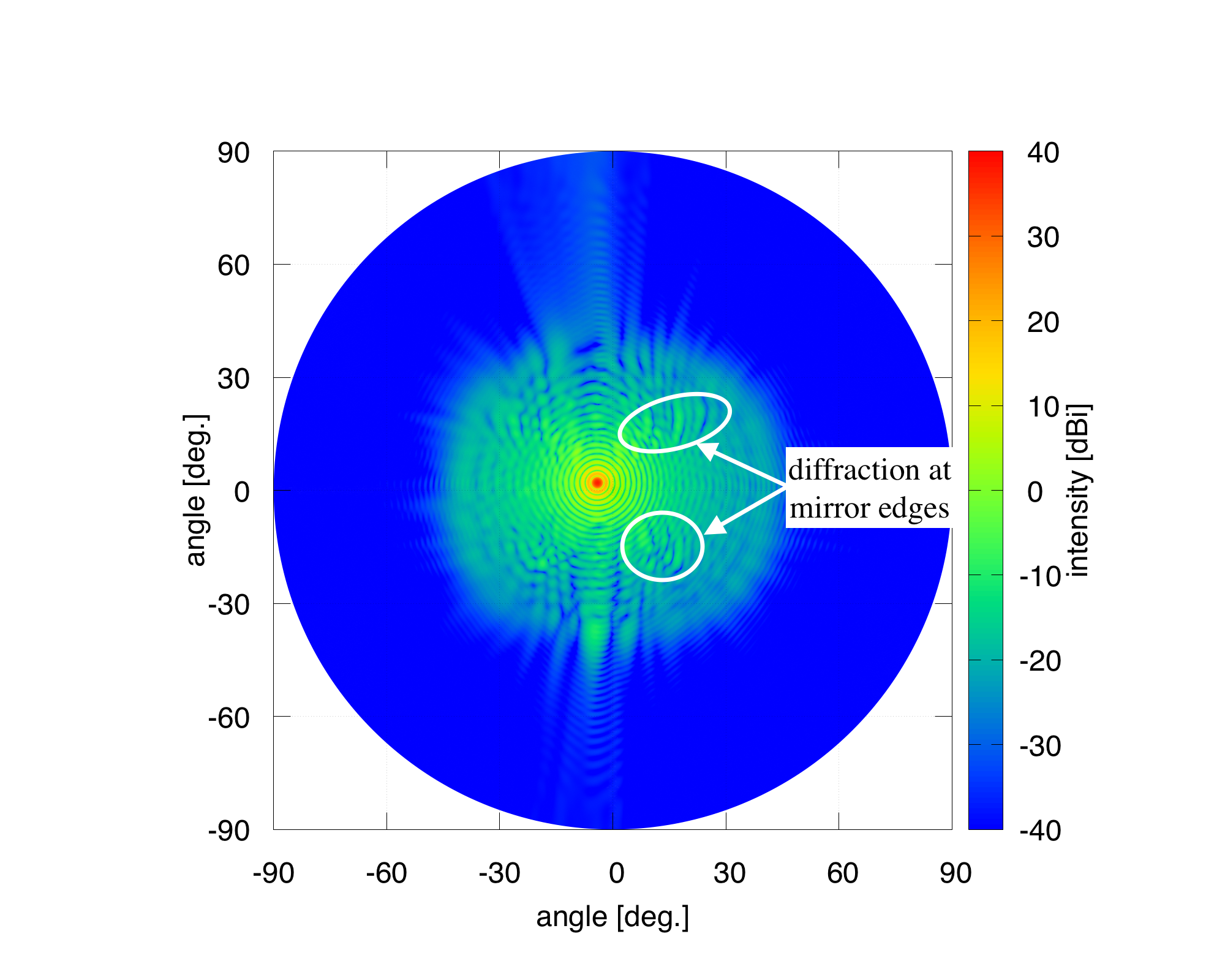} \\
(a)
\end{minipage}
\begin{minipage}{0.5\hsize}
\centering
\includegraphics[height = 7cm]{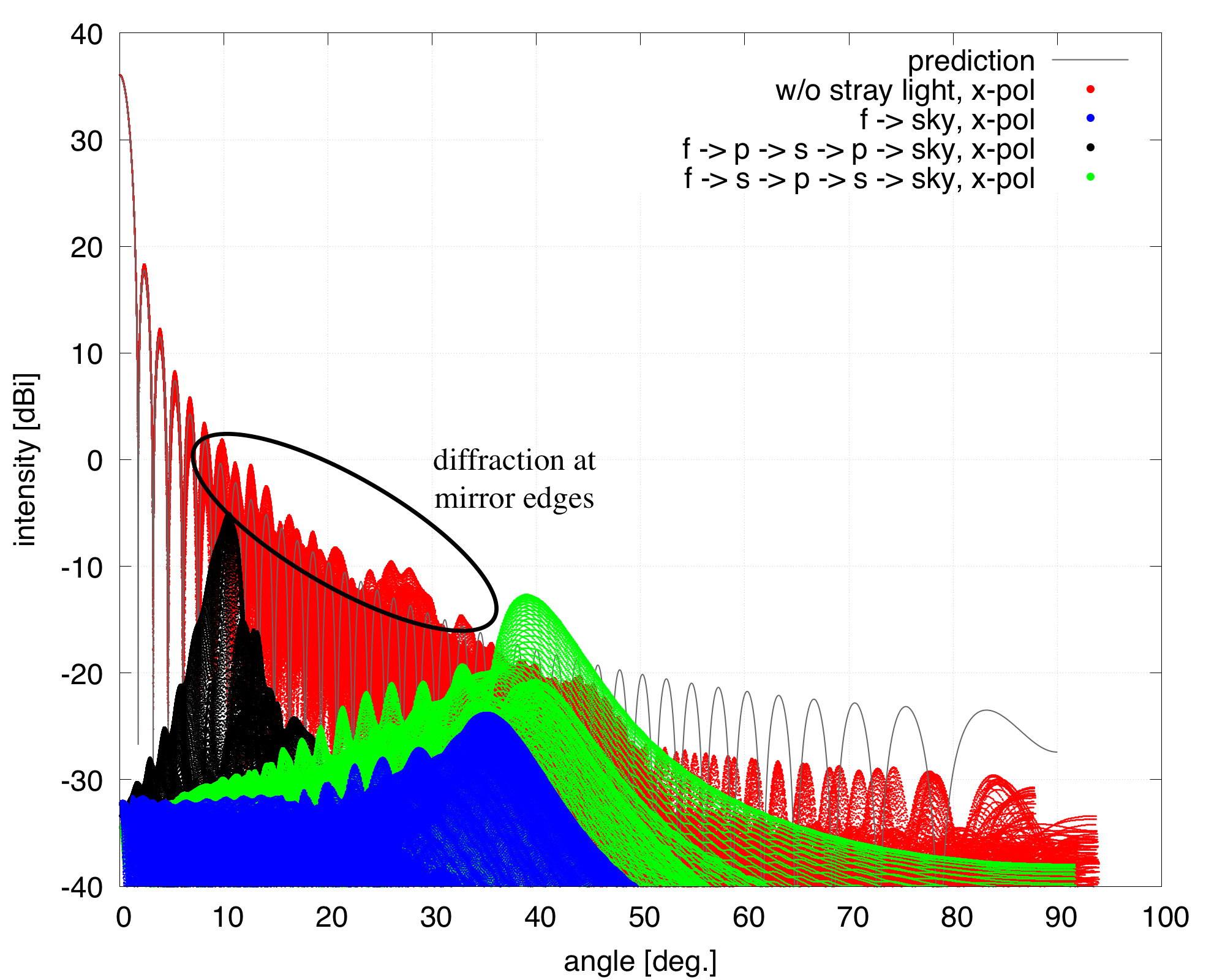} \\
(b)
\end{minipage}
\caption{Physical optical simulation for 30\,GHz, which is out of the band. (a) 2D map. The features from the diffraction at the mirror edges can be found at the right side of the main lobe. (b) 1D cut.}
\label{fig:30GHz_result}
\end{figure}

We have simulated the antenna pattern at $30\,$GHz, as shown in Figure \ref{fig:30GHz_result}, since a bandpass filter cannot cut off sharply at a specific frequency, e.g., 34$\,$GHz, which causes a red leak to the sidelobe. 
The feed here is polarized along the $x$ axis, and located at ($x, y$) = ($-88\,$mm, $+44\,$mm) with a diameter of $24\,$mm, which is different from the current design, but the qualitative effects are the same.
Several features, originating from the diffraction at the mirror edges, are shown within circles in both panels.
These features are at a higher level than that of the nominal diffracted point spread function (PSF). 

The current simulations take into account the reflectors, the aperture stop and the front baffle with perfect absorbers. 
The followings items will be considered for further studies, which might generate additional side-lobes.
\begin{itemize}
\item Actual absorbers have finite reflections on the aperture stop, front hood, detector hood, frame, and panels. 
The absorbers covering the optical cavity and the focal plane are not ideal and they have frequency dependence as well as angle dependence of reflectance.
\item There are multiple reflections (i.e. ghost effects) or multiple scattering among the HWP, the focal plane, the aperture stop, quasi-optical LP Filters, and the absorbers.
\end{itemize}



\subsection{Other optical components}

The aperture stop at 4.8 K with an inner diameter of 400 mm is made of millimeter absorber, TK-RAM\cite{TK-RAM,Saily2004} on an aluminum plate.
This works to make good beam shape for a relatively low edge taper of about $3\,$dB configuration.

Millimeter absorbers to reduce reflections are attached on the inside surface of the $5\,$K frame, which plays a role of a cavity.
Eccosorb AN72 and HR10 are candidates for such absorbers; however, 
they have large TML (total mass loss) and CVCM (collected volatile condensable materials).
According to the NASA outgass database\cite{NASAoutgass}, AN72 washed with ethanol shows reasonable TML and CVCM.

The front-hood, as shown in Figure \ref{fig:lft-structure}, is made of millimeter absorber Eccosorb AN72 and aluminum plate. 

\subsection{Thermal control}

Temperature stability of the optical components of LFT is required to meet the specification of the single detector $f_{\rm knee} = $ $20\,$mHz, which corresponds to $50\,$seconds.
The noise equivalent temperature (NET) of each detector is around $50 \mu$K/$\sqrt{\rm Hz}$, so the noise is integrated to $\Delta T = 7 \, \mu$K in the $50\,$seconds.
It is necessary to meet the following constraint:
\begin{equation}
( \Delta T )^2 \gg \sum_{o=1}^{N_{\rm o}} \left (\delta T_o \times \eta_o \times \epsilon_o \times ({\rm optical\, efficiency}) \right)^2,
\end{equation}
where $N_{\rm o}$ is the number of optical components,  $\delta T_o$ is the temperature stability of the optical components, $\eta_o$ is the optical load fraction and,
$\epsilon_o$ is the emissivity of the optical components.
The optical efficiency of the feed is assumed to be 0.69.
The noise contribution of each optical component is assumed less than $2 \,  \mu$K.
The derived specifications on the stability of the LFT optical components are shown in Table \ref{tbl:lft-temp-stability-req}.
Those specifications give a rough estimate for temperature stability of $\delta T_o/T_o \sim 10^{-5}$ in the worst case, but, more accurate estimates are required, because the optical load fraction ($\eta_o$) depends on the focal plane position, the feed sidelobe and the frequency, as described in section \ref{sect:optical-simulation} and in Figure \ref{fig:GRASP34-imada}.

The temperature of the aperture stop, and other optical components, are planned to be stabilized with heaters to reduce the $1/f$ noise level.

\begin{table}
\begin{center}
\caption{specifications for temperature stability  on the scale of 50 seconds of LFT optical components. The optical load fraction ($\eta_o$) is a typical value, because it depends on the focal plane position, the feed sidelobe, and frequency. $\epsilon_o$ is the emissivity of the optical components.}
\begin{tabular}{lccllrl}
\hline
Components & \multicolumn{2}{c}{Temperature [K]}  & $\eta_o$ & $\epsilon_o$ & Stability \\ 
 & min. & max. &  &   & [mK] \\ \hline
Front hood & 5 & 6 & 0.004 & 0.99 & 3 \\
PMU/HWP & 4.5 & 20 & 0.63 & 0.01 & 0.5 \\
PMU mount & 4.5 & 20 & 0.004 & 0.99 & 0.7 \\
Around aperture stop & 4.5 & 4.8 & 0.2 & 0.99 & 0.02 \\
$5\,$K frame & 4.5 & 5 & 0.1 & 0.99 & 0.03 \\
LFT reflectors & 4.5 & 5 & 0.9 & 0.002 & 1.6 \\
Detector hood & 1.8 & 2 & 0.08 & 0.99 & 0.04 \\
Low-pass filter & 1.7 & 2 & 0.9 & 0.01 & 0.3 \\
\hline
\end{tabular}
\label{tbl:lft-temp-stability-req}
\end{center}
\end{table}


\section{Structure design}

\begin{figure}[ht]
\begin{minipage}{0.5\hsize}
  \begin{center}
\includegraphics[width = 0.8\textwidth]{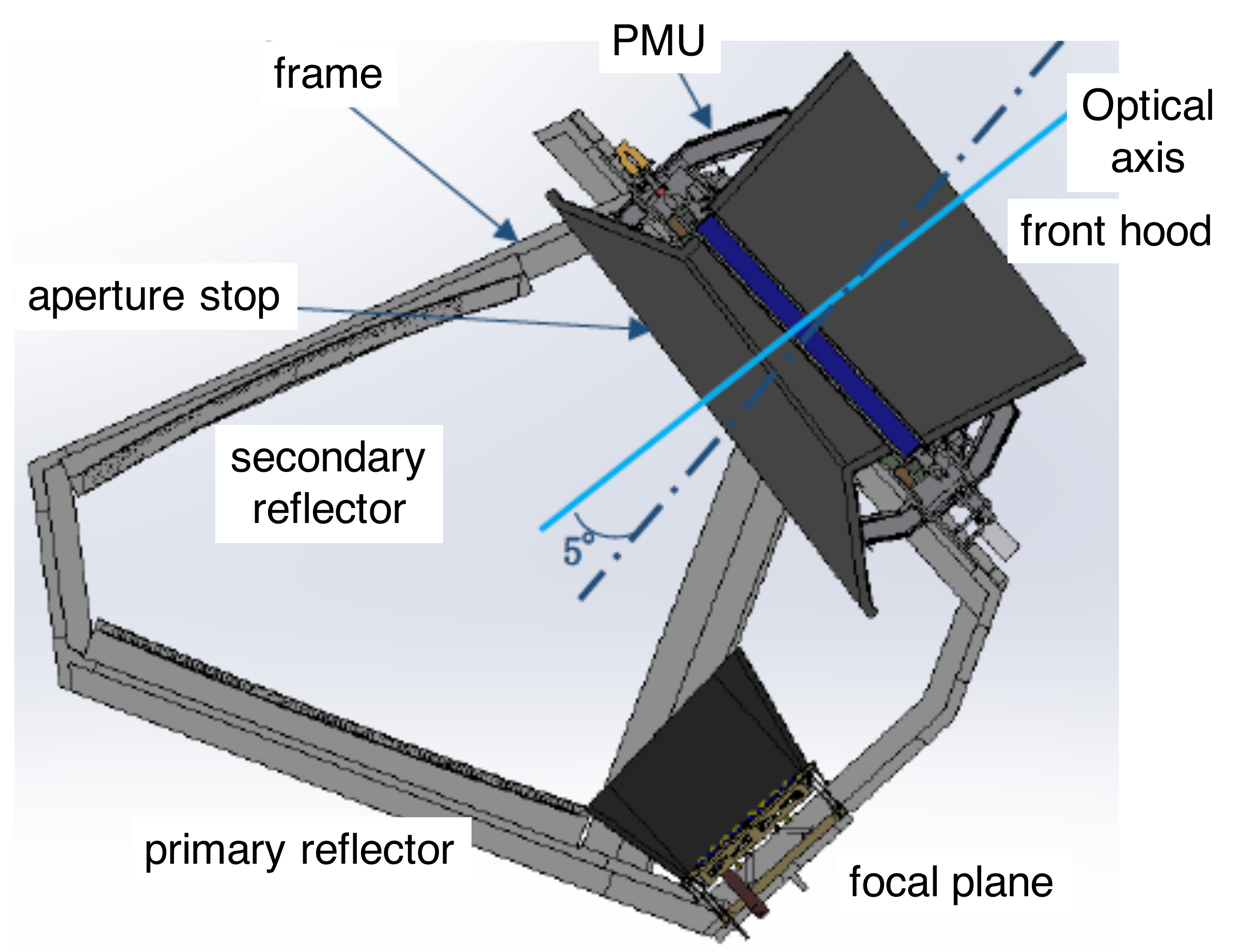}
\end{center}
 \end{minipage}
 \begin{minipage}{0.5\hsize}
  \begin{center}
  \includegraphics[width = 0.8\textwidth]{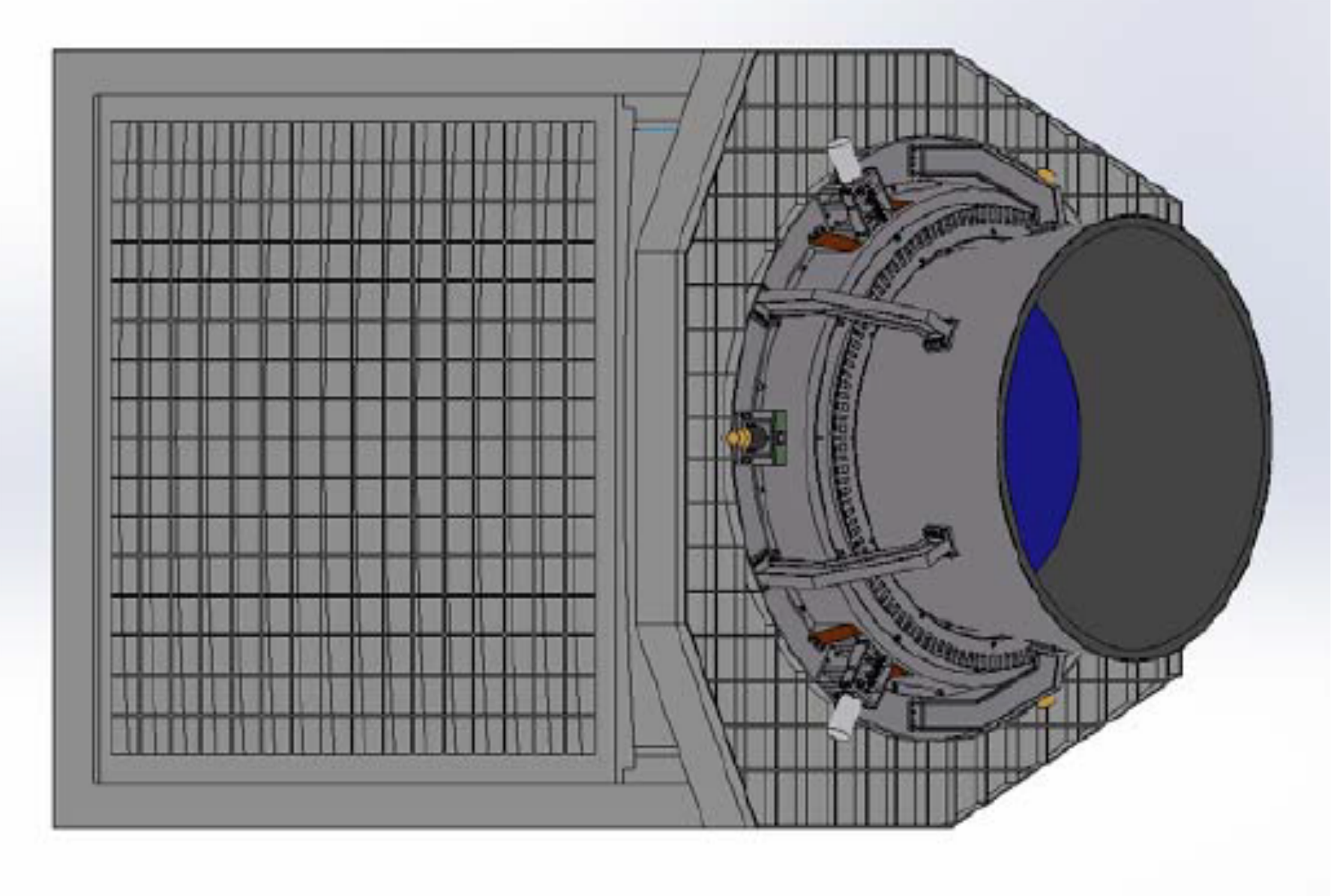}
  \end{center}
 \end{minipage}
 \caption{(Left) Lateral view of structural design of LFT. The side panel is covered with millimeter absorbers.
 (Right) Top view of LFT.}
\label{fig:lft-structure}
\end{figure}

The structural design of LFT is shown in Figure \ref{fig:lft-structure}.
The frame and reflectors of LFT are made of aluminum in order to shrink similarly within 0.4$\,$\% from $300\,$K to $5\,$K\cite{NIST-cryogenics}.
Structural and thermal stability of the telescope is required for the all sky survey of CMB polarization observations.
Aluminum has good thermal conductance at $5\,$K and is mechanically stable.
The frame has structural interfaces at $5\,$K with PMU and the focal plane, which is operated at 0.1 K.
The fastener between the reflector and the frame is planned to use SUS (stainless steel) bolts.
The SUS bolts generate local deformations with an area of several mm, which does not affect on the global shape of the reflectors.
The telescope is supported by trusses made of aluminum on the $5\,$K interface plate.
The total mass of LFT, including the trusses, the PMU and the focal plane, is estimated to be 200 kg.

Optical tolerance analysis leads to  alignment specifications of LFT (Table \ref{tbl:LFT-alignment}), which are derived from the polarization angle variation.
The gravitation deformation of LFT is estimated to be $\delta x$ of $-14\, \mu$m,  $\delta y$ of $-23 \, \mu$m,  $\delta z$ of $22 \, \mu$m, which are all reasonably small. Then, we can plan the ground verification and calibration without directional constraints due to gravitational effects.
According to a scaled model (see Section \ref{sect:scaled}), the alignment can be achieved with careful design and assembly.

\begin{table}
    \centering
    \caption{Alignment specifications of LFT. All values are maxima.}
    \begin{tabular}{|l|l|l|l|l|}
    \hline
        Requirement & Primary (M1) & Secondary (M2) & Frame & Combined \\ \hline
        Mechanical shape error & 15 $\mu$m r.m.s.  &   15 $\mu$m r.m.s. &  &  30 $\mu$m r.m.s.  \\ \hline
        Alignment dx & ± 0.1 mm & ± 0.1 mm & ± 0.2 mm & ± 0.4 mm \\ \hline
        Alignment dy & ± 0.1 mm & ± 0.1 mm & ± 0.2 mm & ± 0.4 mm \\ \hline
        Alignment dz & ± 0.2 mm & ± 0.2 mm & ± 0.2 mm & ± 0.6 mm \\ \hline
        Tilt Rot-x & ± 0.5 arcmin & ± 0.5 arcmin & ± 0.6 arcmin & ± 1.6 arcmin \\ \hline
        Tilt Rot-y & ± 0.4 arcmin & ± 0.4 arcmin & ± 0.2 arcmin & ± 1.0 arcmin \\ \hline
        Tilt Rot-z & ± 0.1 arcmin & ± 0.1 arcmin & ± 0.2 arcmin & ± 0.4 arcmin \\ \hline
    \end{tabular}
\label{tbl:LFT-alignment}
\end{table}

The surface roughness of the reflectors are  designed to be 2--4$\, \mu$m in R$_a$ on the scale of 10 mm, which reduces infrared radiation, mainly from the Galactic plane. 
According to the Ruze fomula $\eta_e = \exp \left[ - \left( \frac{4 \pi \epsilon}{\lambda} \right)^2 \right] $,  infrared radiation more than 5--10$\,$THz (30--60$\, \mu$m) can be scattered.

The telescope is tightly covered with aluminum and absorbers to reduce the stray light from the inner surface of the $30\,$K V-groove (see Figure \ref{fig:LFT-overview}).
The absorber, made of plastic and carbon, is adhered to a panel with epoxy, then the panel is fixed to the $5\,$K frame. 
The cryogenic contraction of the absorber and the epoxy will be carefully designed not to deform the frame. 

\section{LFT Polarization modulation unit (PMU)}

\begin{figure}
\centering
\includegraphics[width = 0.5\textwidth]{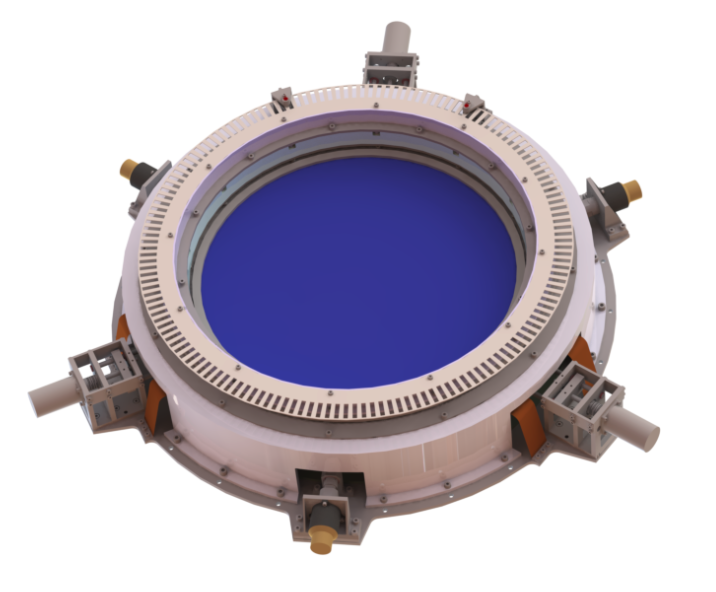}
\caption{LFT Polarization Modulation Unit (PMU)\cite{Sakurai2020}. The sapphire half-wave plate is shown in blue.}
\label{fig:lft_pmu}
\end{figure}

A polarization modulation unit with a transmissive sapphire HWP has been developed for LiteBIRD (Figure \ref{fig:lft_pmu}) \cite{Sakurai2018,Sakurai2018b,KKomatsu2019}.
The progress of the PMU is separately reported\cite{Sakurai2020}.
The PMU/HWP is placed in front of the aperture stop or entrance pupil of $400\,$mm diameter.
The HWP continuously rotates with $46\,$rpm = $0.77\,$Hz. 
PMU uses superconducting magnets for levitation \cite{Sakurai2018}. 
The eddy current and magnetic hysteresis dissipate and increase the temperature of the rotating HWP from $5\,$K to $20\,$K.
The HWP rotation axis is tilted by 5$^\circ$ with respect to the optical axis
 to mitigate multiple reflections including optical ghosts between the HWP and the focal plane.


We have derived following the interface specifications on LFT PMU and focal plane from the LFT specifications (Section \ref{sct:LFT-requirements}) and system designs during ISAS pre-phase A2 \cite{Hazumi2020,Sekimoto2018}.

\begin{enumerate}
\item The optical effects of the observation frequency of 34--161$\,$GHz due to the PMU are minimized to meet the near and far sidelobes specifications of LFT.
\item The opaque $20\,$K parts of PMU are designed to reduce the optical loading.
\item The mass of PMU is $30\,$kg.
\item The heat loads to the $5\,$K stage including the PMU wire harness are less than $3\,$mW.
\item AC magnetic field variation and DC magnetic field are minimized to reduce the effects on the focal plane.
\end{enumerate}

\section{LFT Focal plane}

\begin{figure}[ht]
\begin{minipage}{0.5\hsize}
  \begin{center}
\includegraphics[width = \textwidth]{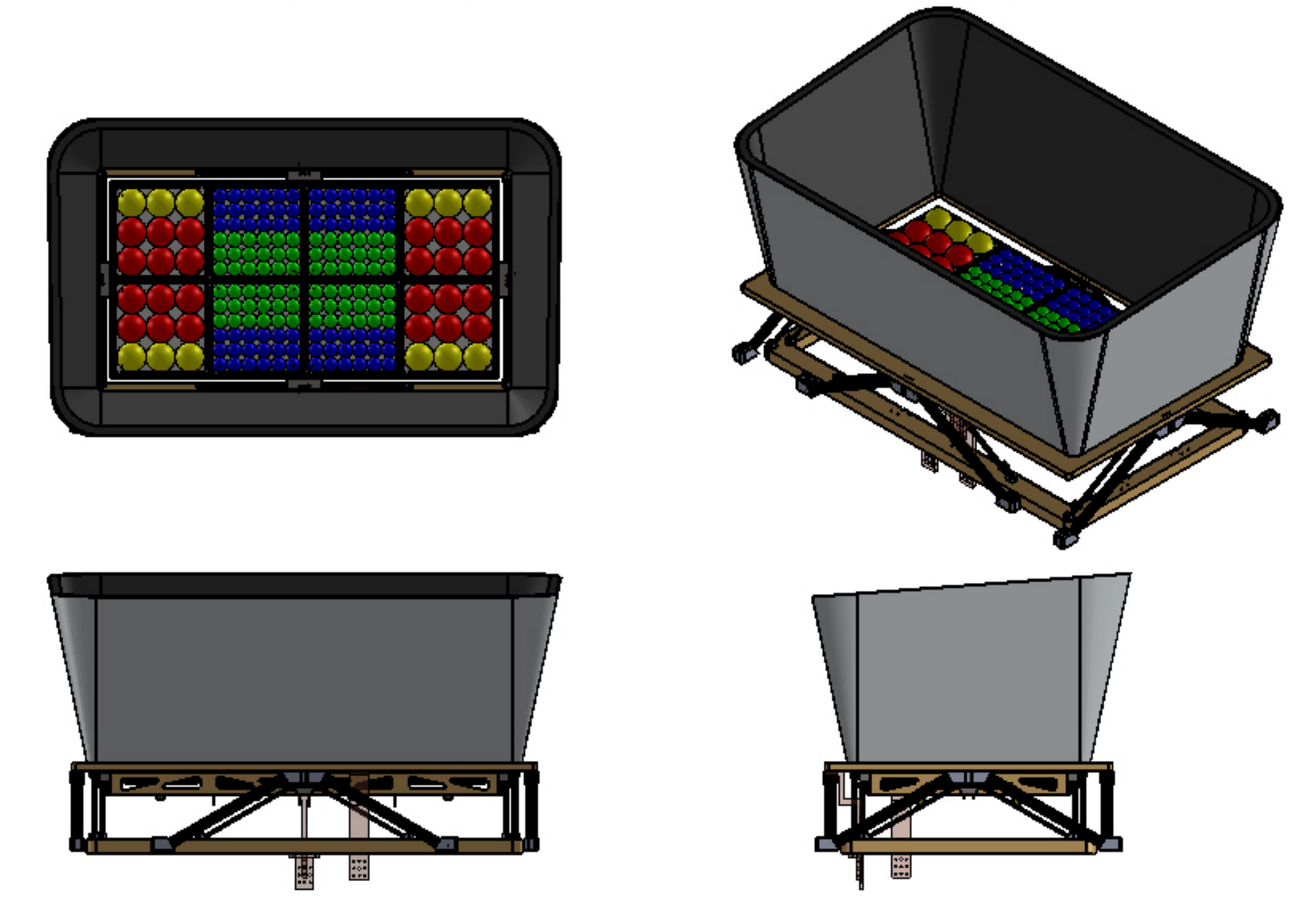}
\end{center}
 \end{minipage}
 \begin{minipage}{0.5\hsize}
  \begin{center}
  \includegraphics[width = 0.95\textwidth]{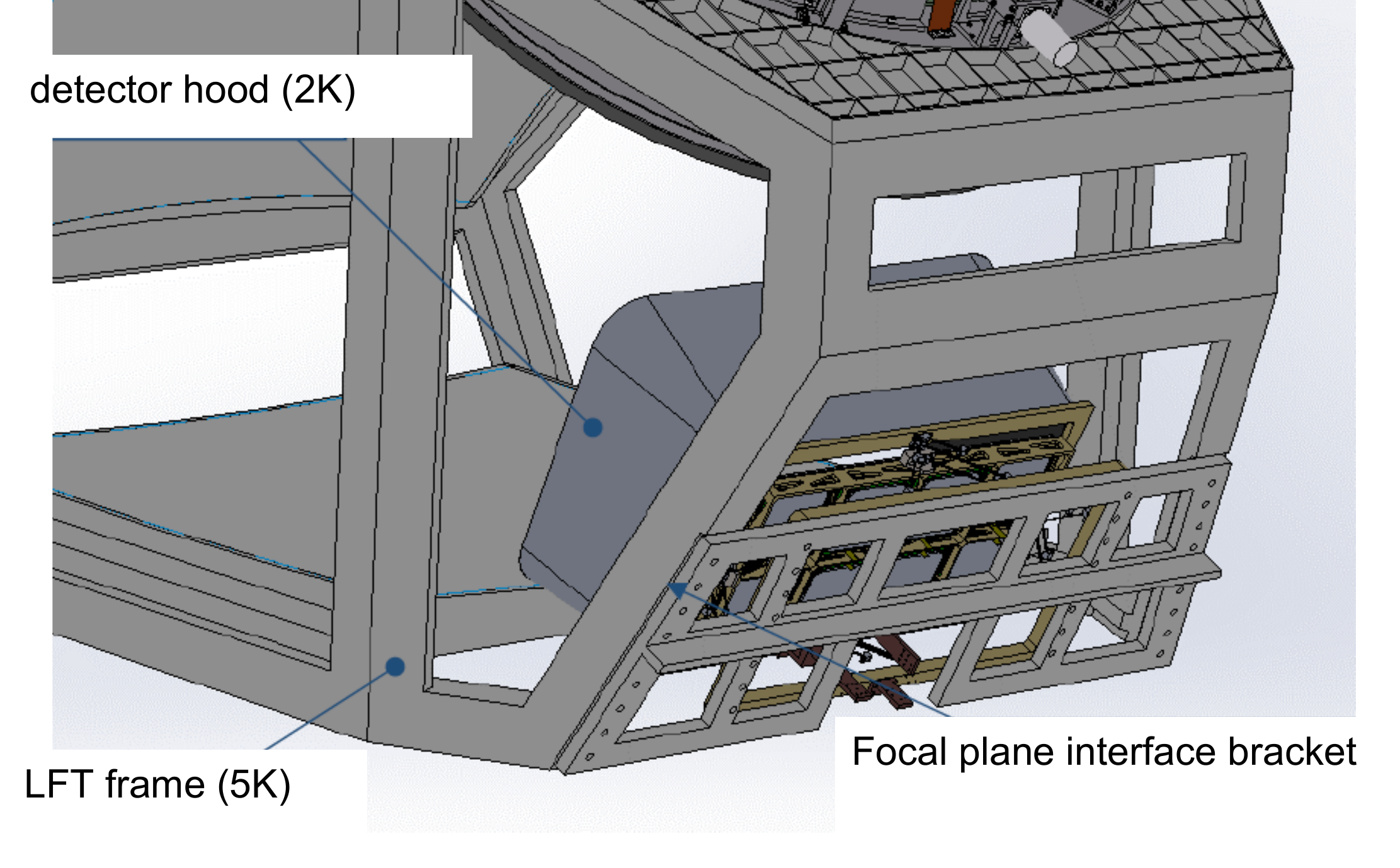}
  \end{center}
 \end{minipage}
 \caption{(Left) LFT focal plane assembly.
 (Right) Structural interface between the focal plane and LFT. }
\label{fig:lft-fp-interface}
\end{figure}

The LFT focal plane has been designed and developed with antenna-coupled TES detectors\cite{Aritoki2018}. 
The lens and sinuous antenna have broadband capability\cite{Edwards2012}.
The focal plane with AlMn TES is cooled to 100 mK with ADRs \cite{Duval2020}.
The cold readout with SQUID amplifiers is also cooled to 100 mK.
Cosmic ray mitigation has been extensively investigated\cite{Stever2020,Tominaga2020}.
The progress of the LFT focal plane is separately reported\cite{Westbrook2020}.

The focal plane consists of eight square (10 cm $\times$ 10 cm) tiles, as shown in Figure \ref{fig:lft-fp-interface}.
The focal plane is shielded with a hood at 2$\,$K to reduce stray light (see Figure \ref{fig:lft-optical}).
A quasi-optical metal-mesh low-pass filter\cite{Ade2006}  is put in front of square modules to reduce thermal loads from far-infrared radiation of the Galactic plane and the $20\,$K radiation of PMU.
A magnetic shield to reduce magnetic variation from the PMU covers the focal plane except for the optical input.
The structural interface at $5\,$K between the focal plane and LFT is designed as shown in Figure~\ref{fig:lft-fp-interface}.

The following interface specifications on the focal plane are flown down from the LFT specifications and system designs.
\begin{enumerate}
\item The optical efficiency of each detector is higher than 0.69.
\item The return loss of the feeds in the in-band frequencies is better than $-10\,$dB.
\item The main beam width of the feeds is consistent with the Gaussian beam radius defined in Table \ref{tbl:feed-parameters} within 5$\,$\%.
\item The sidelobes of each detector are less than $-17\,$dB. Figure  \ref{fig:GRASP34-imada} shows the effects of the feed sidelobes.
\item The optical cross talk among pixels is less than 0.03$\,$\%.
\item The lower frequency edges of 34$\,$GHz and 60$\,$GHz of the 40$\,$GHz band and the 68$\,$GHz band, respectively, have sharper cut-offs to reduce the contamination of sidelobes of the lower frequencies. Figure \ref{fig:30GHz_result} shows the beam pattern at 30 GHz.
\item The polarization efficiency of the feeds should be higher than 98 \%, which corresponds to the cross polarization of $< -17$ dB.
\item The polarization angle of each detector across the frequency band changes by less than $\pm 5^\circ$.
\item The detector noise is basically the photon noise limit of the cosmic microwave background of 2.7$\,$K.
The NET is tabulated in Table \ref{tbl:frequency-bands}.
\item The common mode $1/f$ knee noise of the detector module is stable to be better than 100$\,$mHz.
\item The $1/f$ knee of each detector is stable to be better than 20$\,$mHz.
\item Micro-vibration of the $5\,$K interface is less than 30 $\mu$G/$\sqrt{\rm Hz}$ and 80 $\mu$G/$\sqrt{\rm Hz}$  over 10--200 Hz and 200--500 Hz, respectively.
Under this condition, the focal plane shall perform the required sensitivity.
This requirement is based on the experience of the Hitomi X-ray satellite \cite{Takei2018}.
\item The detector yield including the readout electronics is larger than 80 \%.
\item The dead time fraction due to cosmic ray glitches is less than 0.05.
\item The mass of the focal plane assembly is assumed to be 17 kg without the magnetic shield.
\item The first eigen-frequency of the focal plane is required to larger than 141 Hz for all three axes.
\end{enumerate}

\section{Scaled model demonstration}
\label{sect:scaled}

\begin{figure}[ht]
\begin{minipage}{0.6\hsize}
  \begin{center}
\includegraphics[width =80mm]{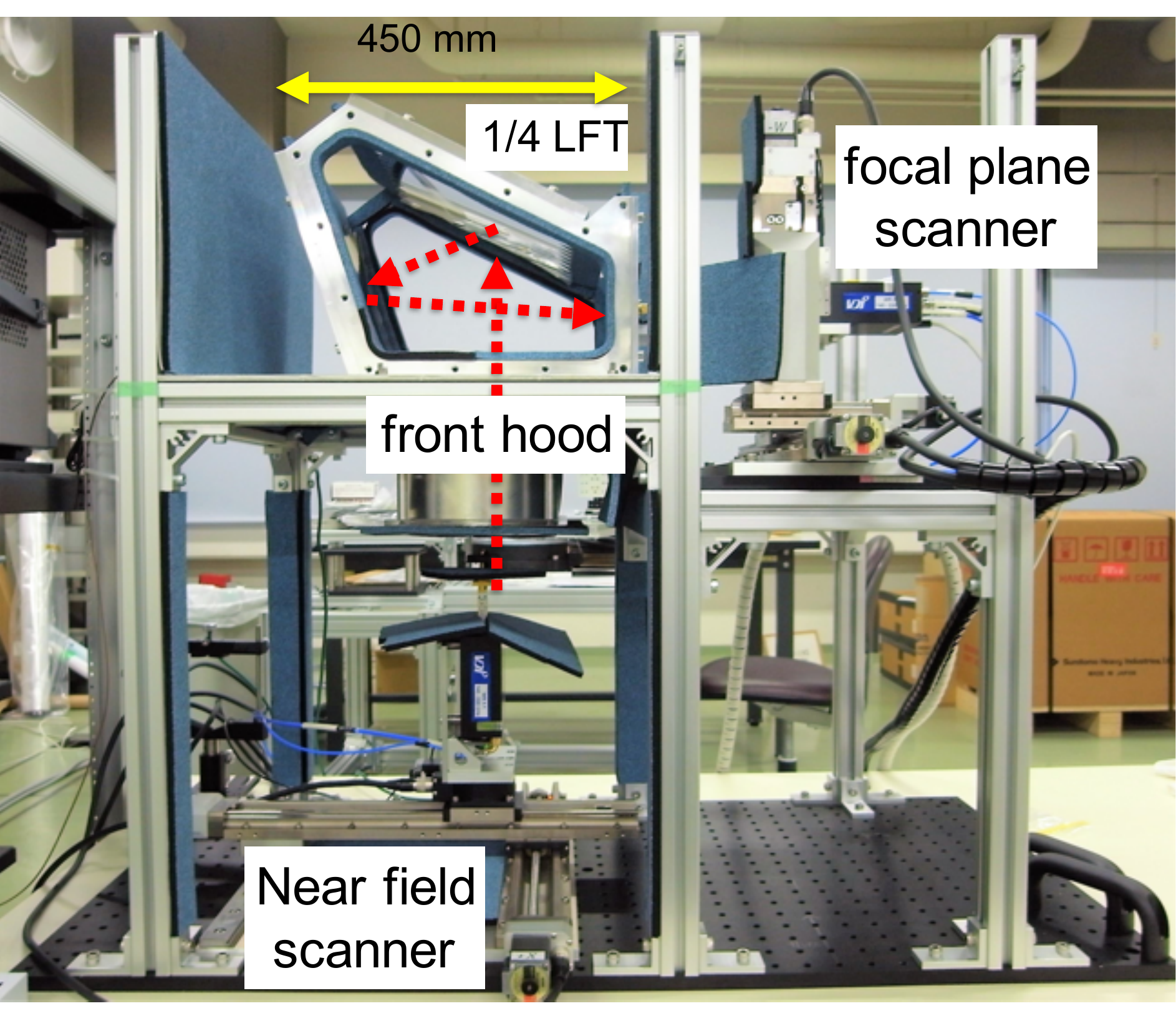}
  \end{center}
 \end{minipage}
 \begin{minipage}{0.4\hsize}
  \begin{center}
     \includegraphics[width=40mm]{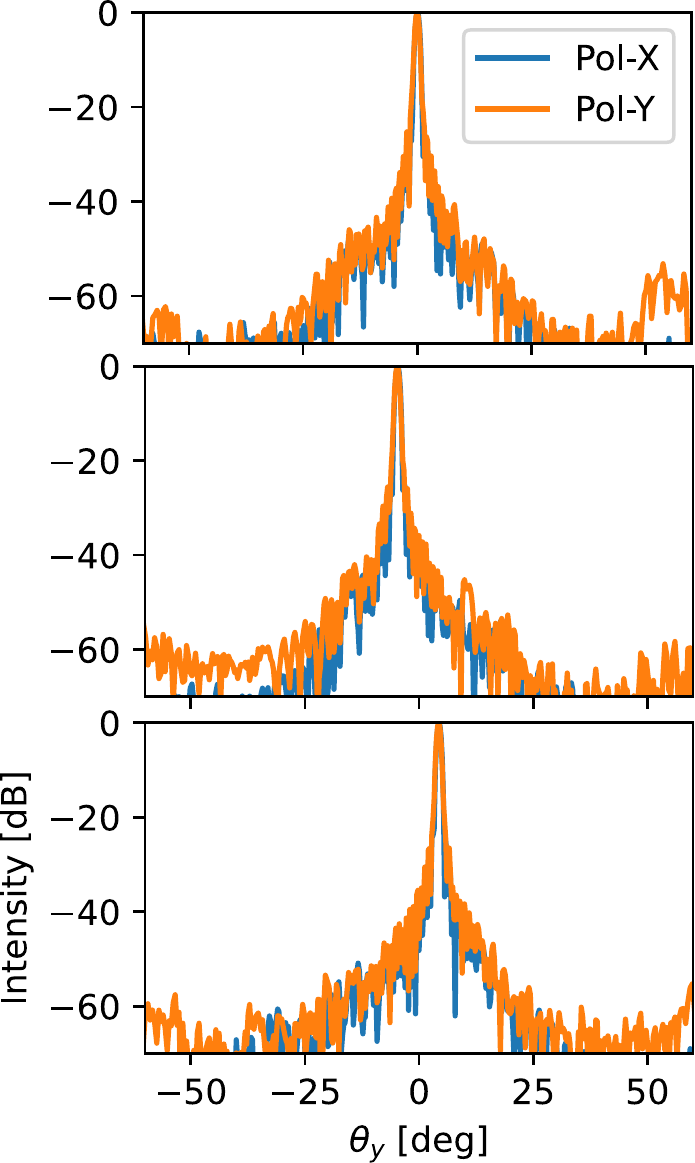}
  \end{center}
 \end{minipage}
 \caption{(Left) LFT quarter (1/4) scaled model and the near-field measurement system \cite{HTakakura2019}. (Right) Far-field patterns of the quarter LFT at the center (top) and edges (middle and bottom) of the focal plane, measured at 220 GHz, which corresponds to 55 GHz in the full model \cite{HTakakura2019}.}
 \label{fig:LFTqscaleMeasSys}
\end{figure}

A quarter (1/4)-size scaled model of the LFT antenna has been designed and developed to 
verify the wide-field design.
Measured frequencies are also scaled, so the antenna pattern of the scaled model reveals that of the full size.

The near-field measurement system with the scaled LFT has been developed as shown in Figure \ref{fig:LFTqscaleMeasSys}\cite{HTakakura2019}.
Measured amplitude and phase data are transformed to far fields.
Figure \ref{fig:LFTqscaleMeasSys} shows far-field beam patterns at three focal positions (see Figure \ref{fig:LFT-focal-plane-pixel}), center, top-right edge, and bottom-right edge, at the frequency of 220 GHz, which corresponds to 55 GHz in the full size LFT.
We confirmed the suppression of far sidelobes based on the scaled model measurements.

Rotation of polarization angle over the field of view is another key parameter for the wide-field design.
A dedicated compact antenna test range (CATR), or a collimated millimeter-wave source has been developed to measure the polarization angle across the wide field of view of the 1/4 LFT.
The polarization angle of the 1/4-scaled LFT has been measured with a resolution of 0.1$'$\cite{HTakakura2020}. 
The polarization angle of polarization $x$ or horizontal polarization was measured to rotate by around 60$'$ across the focal plane, 
while the angle of polarization $y$ or vertical polarization rotates by around 30$'$ across the focal plane.

\begin{figure}[ht]
\begin{minipage}{0.6\hsize}
  \centering
\includegraphics[width =90mm]{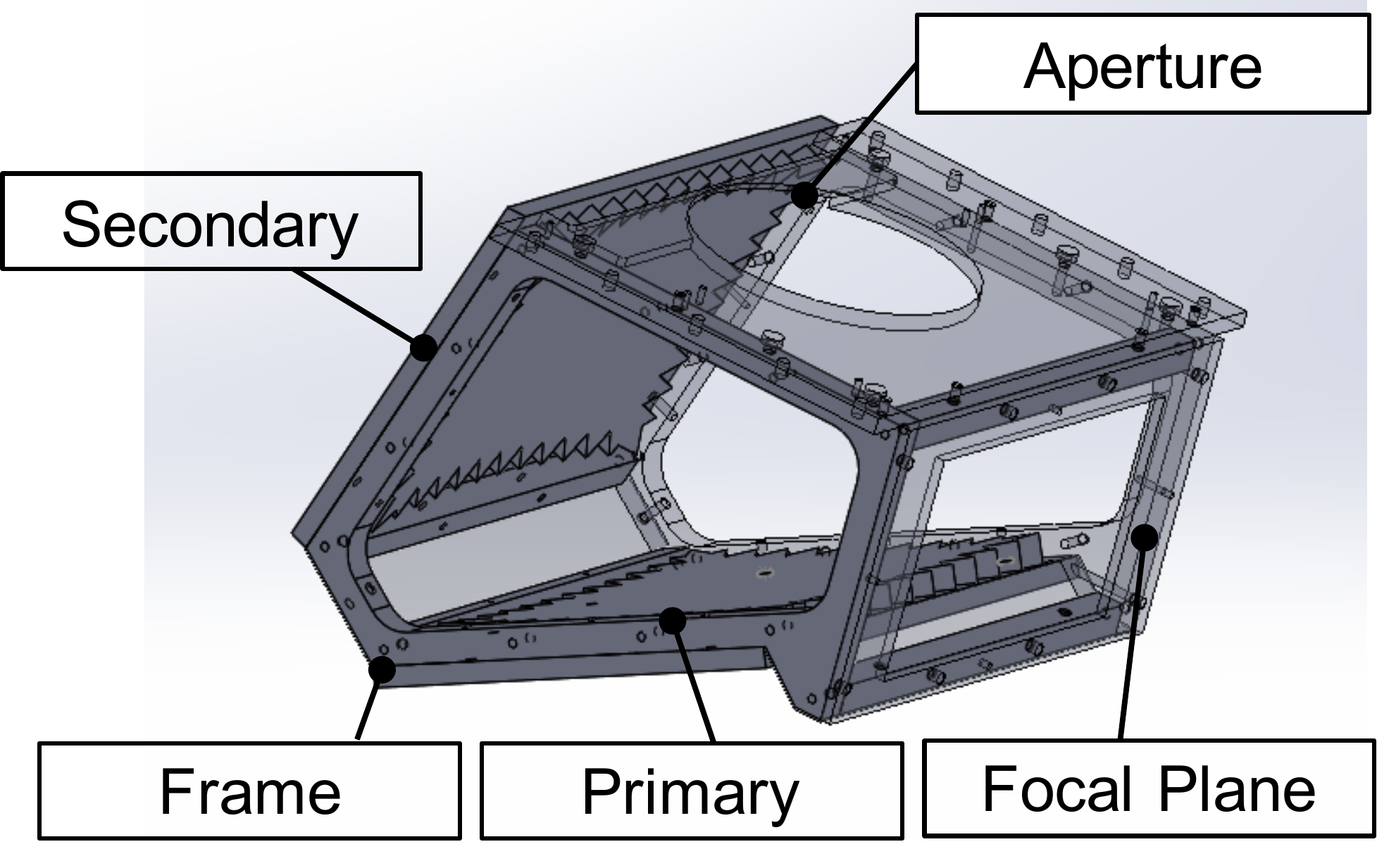}
 \end{minipage}
 \begin{minipage}{0.4\hsize}
 \centering
 \includegraphics[width =50mm]{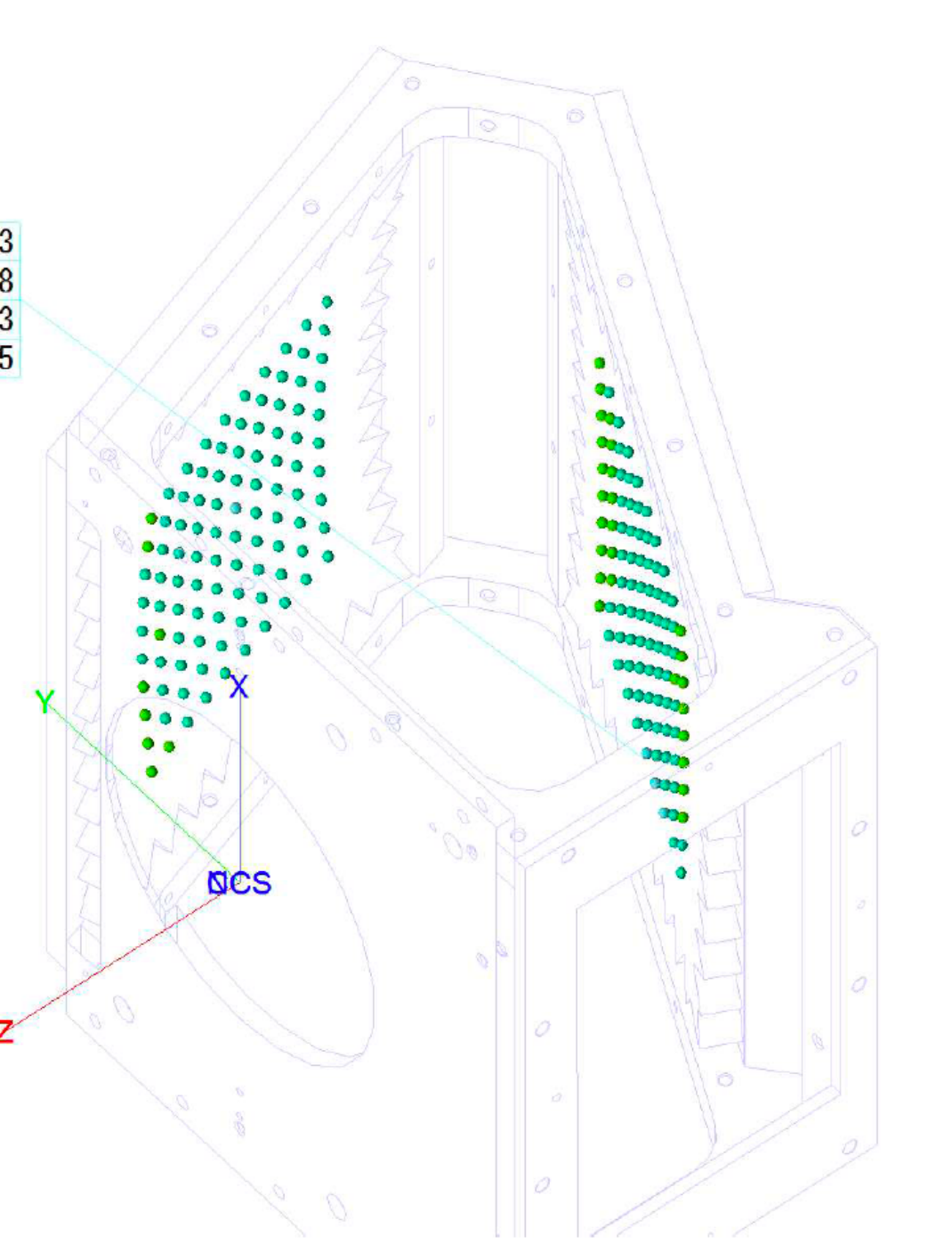}
 \end{minipage}
 \caption{(Left) LFT quarter (1/4) scaled model. (Right) Measurement of reflector surfaces with a coordinated machine.}
 \label{fig:quarterLFT2}
\end{figure}

The structural design of the LFT antenna has been studied with the 1/4-scaled LFT.
The frame structure of the 1/4-LFT as shown in Figure \ref{fig:quarterLFT2}, was assembled with plates and rectangular bars.
The reflector alignment of the assembled 1/4 LFT was measured with a coordinated machine (Mitsutoyo Legex 12128), as shown in Figure \ref{fig:quarterLFT2}.
The fitted curve of the optical surfaces referring to the aperture center is different from  the designed values by $ 36 \, \mu$m and 22$''$ at the maximum.
The measured alignment met the quarter values of the alignment requirement of Table \ref{tbl:LFT-alignment}.

\section{Verification Plan}

Verification and calibration of a cryogenic telescope at the ground facilities before launch are challenging.
A verification plan is tabulated in Table \ref{tbl:verification-plan}.
Two development models (DM/EM and FM\footnote{DM: demonstration model, EM: engineering model, FM: flight model}.) are planned \cite{Sekimoto2018}.

\begin{table} [htb]
    \centering
    \caption{Verification plan of LFT. }
    \begin{tabular}{|l|l|c|c|c|}
    \hline
         &  & DM/EM & FM \\ \hline
       \multicolumn{2}{|l|}{LFT-antenna tests at room temperature }    &  &  \\ \hline
         & Shape measurements with a 3D coordinated machine  & $\checkmark$ & $\checkmark$ \\ \hline
         & Millimeter-wave antenna pattern with horns  & $\checkmark$ & $\checkmark$ \\ \hline
         & V-grooves/MHFT diffraction &  $\checkmark$ & $-$ \\ \hline
       \multicolumn{2}{|l|}{LFT-antenna cryogenic tests at $5\,$K}    &  &  \\ \hline
         & Strain measurements  & $\checkmark$ & $-$ \\ \hline
         & Deformation measurements: photogrammetry or laser sensing  & optional & optional \\ \hline
         & Millimeter-wave antenna pattern with horns  & optional & optional \\ \hline
           \multicolumn{2}{|l|}{ LFT AIV and calibration with FP and PMU}    &  &  \\ \hline
         & Antenna pattern  & $\checkmark$ & $\checkmark$ \\ \hline
         & Polarization angle   & $\checkmark$ & $\checkmark$ \\ \hline
         & Frequency response  & $\checkmark$ & $\checkmark$ \\ \hline
    \end{tabular}
    \label{tbl:verification-plan}
\end{table}

The antenna pattern of LFT before integration with the focal plane will be tested at room temperature. 
A possible method is a near-field beam measurement \cite{HTakakura2019}  or a CATR measurement \cite{HTakakura2020}. 

Diffraction effects due to V-grooves and structures of MHFT will be evaluated and modeled to be small  enough ($< -60$ dB) as designed at room temperature.
A structure thermal model (STM) of the mission payload is constructed and tested with mechanical coolers to verify structural and thermal performance \cite{Sekimoto2018}.
It will be used to measure the electromagnetic effects of V-grooves at room temperature.

Then, the cryogenic deformation of LFT will be measured to be small enough, as designed.
There are a few methods to measure cryogenic deformation of LFT: 1) strain measurements with strain gauges; 2) photogrammetric measurements; and 3) laser reflection measurements.

To verify the requirements of LFT and to calibrate LFT with the focal plane and the PMU, we have a plan to build a beam measurement system in a cryogenic environment.
There are three methods to measure cryogenic beam patterns, polarization angles and spectral response (Table \ref{tbl:cryogenic-RF-measurements}).
One approach is near-field beam measurements in front of the front hood of LFT. 
To obtain the far-field pattern from the near-field measurements, the phase distribution must be retrieved with a reference source\cite{Smith2007}.

\begin{table}
    \centering
    \caption{Possible cryogenic RF measurements. CATR: compact antenna test range. CW : continuous wave/coherent source.}
    \begin{tabular}{|l|c|c|c|}
    \hline
         & Near Field & CATR with CW & CATR with blackbody \\ \hline
        Phase retrieval & necessary & unnecessary & unnecessary \\ \hline
        Volume & compact & large & large \\ \hline
        Time & longer & fast & faster \\ \hline
        Standing wave & no concern & little concern & no concern \\ \hline
        Pol. angle & difficult & possible & possible \\ \hline
        Spectral response & difficult & possible & $-$ \\ \hline
    \end{tabular}
    \label{tbl:cryogenic-RF-measurements}
\end{table}

\begin{figure}[htb]
\begin{center}
   \includegraphics[width =70mm]{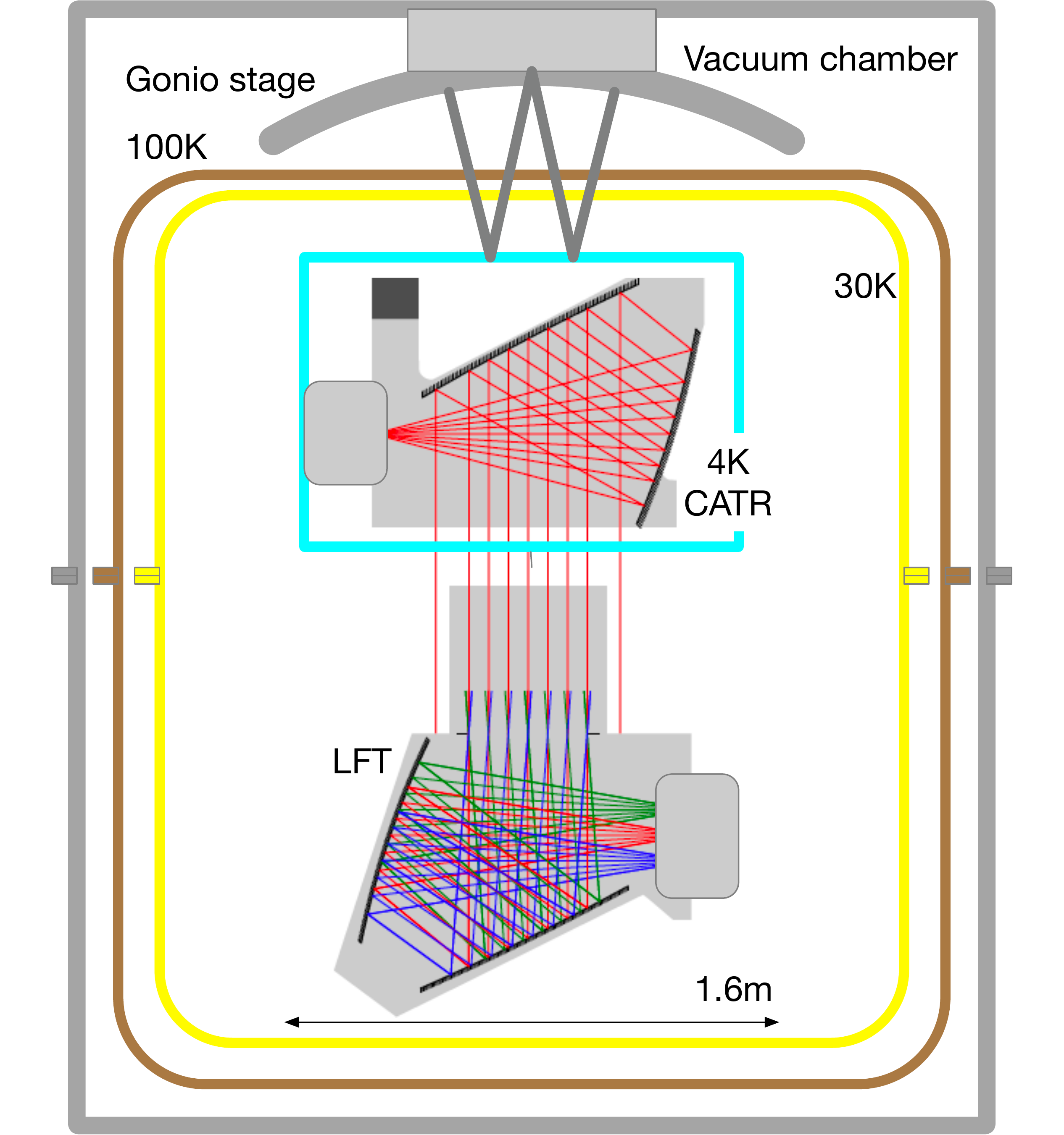}
 \end{center}
 \caption{Schematic drawing of cryogenic set-up with a CATR (compact antenna test range), which moves three-dimensionally with two Gonio stages. It is planned to measure co-polar and cross-polar beam pattens, polarization angle, and spectral response of LFT with CW and blackbody sources.}
 \label{fig:cryo-catr}
\end{figure}

Another method is direct measurement of far-field pattern with a collimated source or a compact antenna test range (CATR), which needs larger volume for the cryogenic environment, as shown in Figure~\ref{fig:cryo-catr}.
This concept has three merits over the phase retrieval near-field beam measurement.
\begin{enumerate}
\item The polarization angle of LFT is also measured with a collimated beam, as demonstrated by H. Takakura et al. 2020\cite{HTakakura2020}.
\item The frequency spectral response is measured with a broadband coherent source. 
A few broadband photo-mixers have been demonstrated at millimeter-wave frequencies \cite{Hirata2002,Kiuchi2017}.
\item It is possible to measure beam patterns with continuum sources as well as coherent sources. Beam measurements with a continuum source are faster than those of multiple frequencies with coherent sources.
\end{enumerate}

In either method, it is crucial to de-couple the mechanics at room temperature from the sources at cryogenic temperature, or to develop moving mechanics operated at low temperature.

\section{Summary}
Based on the performance specifications of LFT, a wide field-of-view design has been studied as well as structural and thermal designs. 
A 1/4-scaled model of LFT has been developed to verify the design. The measured beam pattern was consistent with the optical model at a level of $-50$ dB. 
Interface specifications of the LFT PMU and LFT focal plane are presented.
The verification scheme of LFT is planned as the ISAS/JAXA pre-phase A activity.


\acknowledgments     
 
This work is supported in Japan by ISAS/JAXA for Pre-Phase A2 studies, by the acceleration program of JAXA research and development directorate, by the World Premier International Research Center Initiative (WPI) of MEXT, by the JSPS Core-to-Core Program of A. Advanced Research Networks, and by JSPS KAKENHI Grant Numbers JP15H05891, JP17H01115, and JP17H01125. The Italian LiteBIRD phase A contribution is supported by the Italian Space Agency (ASI Grants No. 2020-9-HH.0 and 2016-24-H.1-2018), the National Institute for Nuclear Physics (INFN) and the National Institute for Astrophysics (INAF). The French LiteBIRD phase A contribution is supported by the Centre National d’Etudes Spatiale (CNES), by the Centre National de la Recherche Scientifique (CNRS), and by the Commissariat \`a l’Energie Atomique (CEA). The Canadian contribution is supported by the Canadian Space Agency. The US contribution is supported by NASA grant no. 80NSSC18K0132. 
Norwegian participation in LiteBIRD is supported by the Research Council of Norway (Grant No. 263011). The Spanish LiteBIRD phase A contribution is supported by the Spanish Agencia Estatal de Investigaci\'on (AEI), project refs. PID2019-110610RB-C21 and AYA2017-84185-P. Funds that support the Swedish contributions come from the Swedish National Space Agency (SNSA/Rymdstyrelsen) and the Swedish Research Council (Reg. no. 2019-03959). The German participation in LiteBIRD is supported in part by the Excellence Cluster ORIGINS, which is funded by the Deutsche Forschungsgemeinschaft (DFG, German Research Foundation) under Germany’s Excellence Strategy (Grant No. EXC-2094 - 390783311). This research used resources of the Central Computing System owned and operated by the Computing Research Center at KEK, as well as resources of the National Energy Research Scientific Computing Center, a DOE Office of Science User Facility supported by the Office of Science of the U.S. Department of Energy.


\bibliography{spie-lb}   

\begin{thebibliography}{10}

\bibitem{Hazumi2012}
M.~Hazumi and {LiteBIRD collaboration}, ``{LiteBIRD: a small satellite for the
  study of B-mode polarization and inflation from cosmic background radiation
  detection},'' in {\em SPIE Astronomical Telescopes + Instrumentation},
  pp.~844219--9, 2012.

\bibitem{Sekimoto2018}
Y.~Sekimoto and {LiteBIRD collaboration}, ``{Concept design of the LiteBIRD
  satellite for CMB B-mode polarization},'' in {\em Space Telescopes and
  Instrumentation 2018: Optical, Infrared, and Millimeter Wave},   {\bf 10698},
  p.~106981Y, SPIE, 2018.

\bibitem{Hazumi2020}
M.~Hazumi and {LiteBIRD Joint Study Group}, ``{LiteBIRD satellite: JAXA's new
  strategic L-class mission for all-sky surveys of cosmic microwave background
  polarization},'' in {\em Space Telescopes and Instrumentation 2020},  SPIE,
  2020.

\bibitem{PhysRevLett.78.2054}
U.~Seljak and M.~Zaldarriaga, ``Signature of gravity waves in the polarization
  of the microwave background,'' {\em Phys. Rev. Lett.}~{\bf 78},
  pp.~2054--2057, Mar 1997.

\bibitem{PhysRevLett.78.2058}
M.~Kamionkowski, A.~Stebbins, A.~Kosowsky, and A.~Stebbins, ``{A Probe of
  Primordial Gravity Waves and Vorticity},'' {\em Phys. Rev. Lett.}~{\bf 78},
  pp.~2058--2061, mar 1997.

\bibitem{PhysRevD.55.1830}
M.~Zaldarriaga and U.~Seljak, ``{All-sky analysis of polarization in the
  microwave background},'' {\em Phys. Rev. D}~{\bf 55}(4), pp.~1830--1840,
  1997.

\bibitem{PhysRevD.55.7368}
M.~Kamionkowski, A.~Kosowsky, and A.~Stebbins, ``{Statistics of cosmic
  microwave background polarization},'' {\em Phys. Rev. D}~{\bf 55}(12),
  pp.~7368--7388, 1997.

\bibitem{Kamionkowski2016}
M.~Kamionkowski and E.~D. Kovetz, ``{The Quest for B Modes from Inflationary
  Gravitational Waves},'' {\em Annu. Rev. Astron. Astrophys.}~{\bf 54},
  pp.~227--69, oct 2016.

\bibitem{Tristram2020}
M.~Tristram, A.~J. Banday, K.~M. G{\'{o}}rski, R.~Keskitalo, C.~R. Lawrence,
  K.~J. Andersen, R.~B. Barreiro, J.~Borrill, H.~K. Eriksen,
  R.~Fernandez-Cobos, T.~S. Kisner, E.~Mart{\'{i}}nez-Gonz{\'{a}}lez,
  B.~Partridge, D.~Scott, T.~L. Svalheim, H.~Thommesen, and I.~K. Wehus,
  ``{Planck constraints on the tensor-to-scalar ratio}.'' {arXiv:
  2010.01139v1}, 2020.

\bibitem{Ludo2020}
L.~Montier and {LiteBIRD Joint Study Group}, ``{Overview of the Mid- and
  High-Frequency Telescopes of the LiteBIRD satellite mission},'' in {\em Space
  Telescopes and Instrumentation 2020},  SPIE, 2020.

\bibitem{Lamagna2020}
L.~Lamagna {\em et~al.}, ``{The optical design of the Litebird Middle and High
  Frequency Telescope},'' in {\em Space Telescopes and Instrumentation 2020},
  SPIE, 2020.

\bibitem{Aritoki2018}
A.~Suzuki {\em et~al.}, ``{The LiteBIRD Satellite Mission - Sub-Kelvin
  Instrument},'' {\em J. of Low Temperature Physics}~{\bf 193}, pp.~1048--1056,
  2018.

\bibitem{Westbrook2020}
B.~Westbrook {\em et~al.}, ``{Detector fabrication development for the LiteBIRD
  satellite mission},'' in {\em Space Telescopes and Instrumentation 2020},
  SPIE, 2020.

\bibitem{Hasebe2019}
T.~Hasebe, Y.~Sekimoto, T.~Dotani, K.~Mitsuda, K.~Shinozaki, and S.~Yoshida,
  ``{Optimization of cryogenic architecture for LiteBIRD satellite using
  radiative cooling},'' {\em Journal of Astronomical Telescopes, Instruments,
  and Systems}~{\bf 5}(4), p.~044002, 2019.

\bibitem{Duval2020}
J.-M. Duval, T.~Prouv{\'{e}}, P.~Shirron, K.~Shinozaki, Y.~Sekimoto, T.~Hasebe,
  G.~Vermeulen, J.~Andr{\'{e}}, M.~Hasumi, L.~Montier, and B.~Mot, ``{LiteBIRD
  Cryogenic Chain: 100 mK Cooling with Mechanical Coolers and ADRs},'' {\em
  Journal of Low Temperature Physics}~{\bf 199}, pp.~730--736, 2020.

\bibitem{Sakurai2020}
Y.~Sakurai {\em et~al.}, ``{Breadboard model of polarization modulator unit
  based on a continuous rotating half-wave plate for low frequency telescope of
  LiteBIRD space mission},'' in {\em Space Telescopes and Instrumentation
  2020},  SPIE, 2020.

\bibitem{Ade2006}
P.~A.~R. Ade, G.~Pisano, C.~Tucker, and S.~Weaver, ``{A review of metal mesh
  filters},'' in {\em SPIE},  pp.~62750U--62750U, 2006.

\bibitem{Ishino2016}
H.~Ishino and {LiteBIRD collaboration}, ``{LiteBIRD: lite satellite for the
  study of B-mode polarization and inflation from cosmic microwave background
  radiation detection},'' in {\em Proc. SPIE},   {\bf 9904}, p.~99040X, 2016.

\bibitem{Ishino2016b}
H.~Ishino, R.~Nagata, and {LiteBIRD working group}, ``{Development of LiteBIRD
  analysis pipeline and systematics evaluation},'' in {\em Proceedings of the
  16th Space Science Symposium},  2016.

\bibitem{Nagata2019}
R.~Nagata, ``{Requirement analysis for LiteBIRD optical system},'' {\em JAXA
  Supercomputer System Annual Report April 2017-March 2018 JSS2
  Inter-University Research} , feb 2019.

\bibitem{Lee2019}
A.~T. Lee {\em et~al.}, ``{LiteBIRD an all-sky cosmic microwave background
  probe of inflation},'' {\em Bulletin of the American Astronomical
  Society}~{\bf 7}(51), p.~286, 2019.

\bibitem{Tsuji2020}
M.~Tsuji {\em et~al.}, ``{Simulating electromagnetic transfer function from the
  transmission antennae to the sensors vicinity in LiteBIRD},'' in {\em Space
  Telescopes and Instrumentation 2020},  SPIE, 2020.

\bibitem{Tran2008}
H.~Tran, A.~Lee, S.~Hanany, M.~Milligan, and T.~Renbarger, ``{Comparison of the
  crossed and the Gregorian Mizuguchi-Dragone for wide-field millimeter-wave
  astronomy},'' {\em Applied Optics}~{\bf 47}(2), pp.~103--109, 2008.

\bibitem{Bernacki2012}
B.~E. Bernacki, J.~F. Kelly, D.~Sheen, B.~Hatchell, P.~Valdez, J.~Tedeschi,
  T.~Hall, and D.~McMakin, ``{Wide-field-of-view millimeter-wave telescope
  design with ultra-low cross-polarization},'' in {\em SPIE},   {\bf 8362},
  pp.~836207--836211, may 2012.

\bibitem{Young2018}
K.~Young {\em et~al.}, ``{Optical design of PICO: a concept for a space mission
  to probe inflation and cosmic origins},'' in {\em Space Telescopes and
  Instrumentation 2018: Optical, Infrared, and Millimeter Wave},   {\bf 10698},
  pp.~1242 -- 1253, SPIE, 2018.

\bibitem{Tran2010}
H.~Tran {\em et~al.}, ``{Optical Design of the EPIC-IM Crossed Dragone
  Telescope},'' in {\em SPIE},   {\bf 7731}, pp.~77311R--15, 2010.

\bibitem{Kashima2018}
S.~Kashima, M.~Hazumi, H.~Imada, N.~Katayama, T.~Matsumura, Y.~Sekimoto, and
  H.~Sugai, ``{Wide field-of-view crossed Dragone optical system using
  anamorphic aspherical surfaces},'' {\em Appl. Opt.}~{\bf 57}(15),
  pp.~4171--4179, 2018.

\bibitem{GRASP}
``{TICRA}.'' \url{http://www.ticra.com/software/grasp/}.

\bibitem{Imada2018}
H.~Imada, T.~Dotani, T.~Hasebe, M.~Hazumi, J.~Inatani, H.~Ishino, S.~Kashima,
  N.~Katayama, K.~Kimura, T.~Matsumura, R.~Nagata, Y.~Sekimoto, H.~Sugai,
  A.~Suzuki, and S.~Utsunomiya, ``{The optical design and physical optics
  analysis of a cross-Dragonian telescope for LiteBIRD},'' in {\em Space
  Telescopes and Instrumentation 2018: Optical, Infrared, and Millimeter Wave},
    {\bf 10698}, p.~106984K, SPIE, 2018.

\bibitem{HFSS}
``{HFSS}.'' \url{https://www.ansys.com/products/electronics/ansys-hfss}.

\bibitem{Edwards2012}
J.~M. Edwards, R.~O'Brient, A.~T. Lee, and G.~M. Rebeiz, ``{Dual-Polarized
  Sinuous Antennas on Extended Hemispherical Silicon Lenses},'' {\em IEEE
  Transactions on Antennas and Propagation}~{\bf 60}, pp.~4082--4091, sep 2012.

\bibitem{TK-RAM}
``{TK-RAM}.'' \url{http://www.terahertz.co.uk/}.

\bibitem{Saily2004}
J.~Saily and A.~Raisanen, ``{Characterization of Submillimeter Wave Absorbers
  from 200 - 600 GHz},'' {\em International Journal of Infrared and Millimeter
  Waves}~{\bf 5}, p.~71, 2004.

\bibitem{NASAoutgass}
``{NASA outgassing data}.'' \url{https://outgassing.nasa.gov/}.

\bibitem{NIST-cryogenics}
``{NIST Material Properties References}.''
  \url{https://trc.nist.gov/cryogenics/materials/references.htm}.

\bibitem{Sakurai2018}
Y.~Sakurai, T.~Matsumura, T.~Iida, H.~Kanai, N.~Katayama, H.~Imada, H.~Ohsaki,
  Y.~Terao, T.~Shimomura, H.~Sugai, H.~Kataza, R.~Yamamoto, and S.~Utsunomiya,
  ``Design and thermal characteristics of a 400 mm diameter levitating rotor in
  a superconducting magnetic bearing operating below at 10 k for a cmb
  polarization experiment,'' {\em IEEE Transactions on Applied
  Superconductivity}~{\bf 28}, pp.~1--4, June 2018.

\bibitem{Sakurai2018b}
Y.~Sakurai {\em et~al.}, ``{Design and development of a polarization modulator
  unit based on a continuous rotating half-wave plate for LiteBIRD},'' in {\em
  Proc. SPIE},   {\bf 10708}, p.~107080E, 2018.

\bibitem{KKomatsu2019}
K.~Komatsu, T.~Matsumura, H.~Imada, H.~Ishino, N.~Katayama, and Y.~Sakurai,
  ``{Demonstration of the broadband half-wave plate using the nine-layer
  sapphire for the cosmic microwave background polarization experiment},'' {\em
  Journal of Astronomical Telescopes, Instruments, and Systems}~{\bf 5}, p.~1,
  nov 2019.

\bibitem{Stever2020}
S.~Stever {\em et~al.}, ``{Cosmic ray glitch predictions, physical modelling,
  and overall effect on the LiteBIRD space mission},'' {\em JCAP} , p.~in
  preparation, 2020.

\bibitem{Tominaga2020}
M.~Tominaga {\em et~al.}, ``Simulation of cosmic ray effects in the cmb b-mode
  polarization observation satellite litebird,'' in {\em Space Telescopes and
  Instrumentation 2020},  SPIE, 2020.

\bibitem{Takei2018}
Y.~Takei, S.~Yasuda, K.~Ishimura, {\em et~al.}, ``{Vibration isolation system
  for cryocoolers of soft x-ray spectrometer on-board ASTRO-H (Hitomi)},'' {\em
  Journal of Astronomical Telescopes, Instruments, and Systems}~{\bf 4},
  p.~011216, feb 2018.

\bibitem{HTakakura2019}
H.~Takakura, Y.~Sekimoto, J.~Inatani, S.~Kashima, H.~Imada, T.~Hasebe, T.~Kaga,
  Y.~Takeda, and N.~Okada, ``{Far-Sidelobe Antenna Pattern Measurement of
  LiteBIRD Low Frequency Telescope in 1/4 Scale},'' {\em IEEE Transactions on
  Terahertz Science and Technology}~{\bf 9}, pp.~598--605, nov 2019.

\bibitem{HTakakura2020}
H.~Takakura, Y.~Sekimoto, J.~Inatani, S.~Kashima, and M.~Sugimoto,
  ``Polarization angle measurement of litebird low frequency telescope scaled
  model,'' in {\em Space Telescopes and Instrumentation 2020},  SPIE, 2020.

\bibitem{Smith2007}
D.~Smith, M.~Leach, M.~Elsdon, and S.~Foti, ``{Indirect Holographic Techniques
  for Determining Antenna Radiation Characteristics and Imaging Aperture
  Fields},'' {\em IEEE Antennas and Propagation Magazine}~{\bf 49}, pp.~54--67,
  feb 2007.

\bibitem{Hirata2002}
A.~Hirata, T.~Nagatsuma, R.~Yano, H.~Ito, T.~Furuta, Y.~Hirota, T.~Ishibashi,
  H.~Matsuo, A.~Ueda, T.~Noguchi, Y.~Sekimoto, M.~Ishiguro, and S.~Matsuura,
  ``{Output power measurement of photonic millimeter-wave and
  sub-millimeter-wave emitter at 100-800 GHz},'' {\em Electron Lett}~{\bf 38},
  p.~798, 2002.

\bibitem{Kiuchi2017}
H.~Kiuchi, T.~Kawanishi, and A.~Kanno, ``Wide frequency range optical
  synthesizer with high-frequency resolution,'' {\em IEEE Photonics Technology
  Letters}~{\bf 29}, pp.~78--81, Jan 2017.

\end{thebibliography}
\bibliographystyle{spiebib}   

\end{document}